\documentclass[]{aa}
\usepackage{graphicx}
\usepackage[varg]{txfonts}
\usepackage{enumitem}
\usepackage{natbib}   
\bibpunct{(}{)}{;}{a}{}{,}  
\usepackage{amsmath,amssymb}
\usepackage{mathtools}
\usepackage{lscape}
\usepackage{longtable}
\usepackage{enumerate}
\usepackage{setspace}
\usepackage[usenames,dvipsnames]{color}
\usepackage[pagebackref=true]{hyperref}%
\hypersetup{colorlinks=true,linkcolor=blue,citecolor=blue,filecolor=magenta,urlcolor=blue}
\usepackage{float,morefloats} 
\begin{document}


\definecolor{orange}{RGB}{255,127,0}
\title{The formation of UCDs and massive GCs:}
\subtitle{Quasar-like objects to test for a variable stellar initial mass function}
\author{T. Je\v{r}\'abkov\'a \inst{1,2} \fnmsep\thanks{tjerabko@eso.org}
\and P. Kroupa \inst{2,1} \fnmsep\thanks{pavel@astro.uni-bonn.de}
\and J. Dabringhausen \inst{1}
\and M. Hilker \inst{3}
\and K. Bekki \inst{4}
}
\institute{Astronomical Institute, Charles University in Prague, V 
Hole\v{s}ovi\v{c}k\'ach 2, CZ-180 00 Praha 8, Czech Republic
\and
Helmholtz Institut f\"{u}r Strahlen und Kernphysik, Universit\"{a}t Bonn, Nussallee 14–16, 53115 Bonn, Germany
\and
European Southern Observatory, Karl-Schwarzschild-Str. 2, D-85748, Garching bei M\"{u}nchen, Germany
\and
ICRAR,The University of Western Australia 35 Stirling Hwy, Crawley Western Australia 6009, Australia
}
\date{Received 24.05. 2017; accepted 21.08. 2017}
\titlerunning{The formation of UCDs: Testing for a variable stellar IMF}

\abstract{
The stellar initial mass function (IMF) has been described as being invariant, bottom-heavy,
or top-heavy in extremely dense star-burst  conditions.
To provide usable observable diagnostics, we calculate redshift dependent spectral energy
distributions of stellar populations
in extreme star-burst clusters, which are likely to have been the precursors of present day 
massive globular clusters (GCs) and of ultra compact dwarf galaxies (UCDs).
The retention fraction of stellar remnants is taken into account to assess the mass to 
light ratios of the ageing star-burst.
Their redshift dependent photometric properties are calculated as predictions for 
James Webb Space Telescope (JWST) observations. While the present day GCs and UCDs are 
largely degenerate concerning bottom-heavy or top-heavy IMFs, a metallicity- and 
density-dependent top-heavy IMF implies the most massive UCDs, at ages <100 Myr,  to appear as objects
with quasar-like luminosities with a 0.1-10\% variability on a monthly timescale due to 
core collapse supernovae.
 }
 \maketitle

\section{Introduction}
The question of whether the stellar initial mass function (IMF) varies systematically with the physical conditions is a central problem of modern 
astrophysics \citep{Elmegreen2004,Bastian2010, Kroupa2013}. 
If it does vary, we would expect the largest differences compared to nearby star forming regions in the most extreme star-burst 
conditions \citep[e.g.][]{Larsen1998}.
Where are the most extreme star-burst conditions in terms of star formation rate densities to be found, with the constraint 
that their present-day remnants are observable allowing detailed observational scrutiny of their present-day stellar populations? 
Given the present-day masses ($>10^5\, M_{\odot}$) and present-day half-light radii ($1-50\, \mathrm{pc}$), globular clusters (GCs) 
and in particular ultra compact dwarf galaxies (UCDs) are promising candidates \citep[][present-day half mass radius]{Mieske2002,Chilingarian2008,Brodie2011}. 
For example a $10^7\, M_{\odot}$ present-day UCD with a half-light radius of $20\, \mathrm{pc}$ would likely have had a  mass of $10^8\, M_{\odot}$ and a radius of a few pc when it was 1 Myr old \citep{Dabringhausen2010}.

Ultra compact dwarf galaxies (UCDs) were discovered by
\cite{Hilker1999} and \cite{Drinkwater2000}.
There is no unique definition of what a UCD is, but it is generally taken to mean a 
stellar system with a radius between a few and $100\,\mathrm{pc}$, and a dynamical mass of 
$10^6\,\mathrm{M_{\odot}}\lessapprox M_{dyn} \lessapprox 10^8\,\mathrm{M_{\odot}}$. Absolute magnitudes 
of UCDs lie roughly between $-10 \gtrapprox M_V \gtrapprox -16\, \mathrm{M_V}$, where the higher value corresponds to an old 
and low mass UCD (similar to $\omega$ Cen), while the lower value is more characteristic for a fairly 
young and/or massive UCDs (like W3). 
The most massive known old UCDs have magnitudes of $M_V \approx -14$.
The presence of UCDs has been reported also in
other galaxy clusters, for example\ Virgo
\citep{Drinkwater2004}, Coma \citep{Price2009}, Centaurus
\citep{Mieske2007}, Hydra I \citep{Misgeld2011}, and Perseus \citep{Penny2012},
or in massive intermediate-redshift clusters \citep[e.g.][]{Zhang2017}.
The existence of systems with properties
of UCDs was discussed already by \cite{Kroupa1998} based on the
observations of star cluster complexes in the Antennae galaxies.

\subsection{UCD formation and the categorization issue} 
The origin and evolution of UCDs is still a matter of debate 
\citep[e.g.][]{Cote2006,Hilker2009,Brodie2011} and up until today several
possible formation scenarios have been proposed: (A) UCDs are the
massive end of the distribution of GCs  \citep[e.g.][]{Mieske2002, Forbes2008, Murray2009,
Dabringhausen2009, Chiboucas2011, Mieske2012, Renaud2015}, (B) UCDs
are merged star cluster complexes
\citep{Kroupa1998, Fellhauer2002a,Fellhauer2002b,Bruens2011}, (C) UCDs are the
tidally stripped nuclei of dwarf galaxies
\citep{Oh1995,Bekki2001,Bekki2003,Drinkwater2003,Goerdt2008,Pfeffer2013}, and  (D) UCDs are
remnants of primordial compact galaxies \citep{Drinkwater2004}.

\cite{Murray2009} showed that the distribution and kinematics of the visible matter is not
consistent with UCDs being objects dominated by non-baryonic dark
matter halos. This is supported by a detailed study of one of
the most massive UCDs \citep{Frank2011}.
Based on the
Millenium II cosmological simulation, \cite{Pfeffer2014} show that the formation of UCDs
as tidally stripped nuclei of dwarf galaxies can account only for about
50 per cent of observed objects with mass $>10^7\,M_{\odot}$; for masses $>10^6\,M_{\odot}$ 
this drops to approximately 20 per cent \citep{Pfeffer2016}.
Furthermore, earlier \cite{Thomas2008}
showed that the tidally stripped nucleus scenario fails to reproduce UCDs
located in the outer parts of the  Fornax cluster.  Based on these results 
we focus on
scenarios (A) and (B), which both suggest that UCDs are
the high-mass end of star cluster-like objects, formed most likely during massive
starbursts \citep{Weidner2004, Schulz2015, Schulz2016} at
higher redshift, where compact star-bursts are indeed observed
\citep{Vanzella2017, Glazebrook2017}.

\subsection{Aim of this work}
 In this contribution we aim to quantify how extreme star formation environments may appear at high redshifts,
where the most intense star-bursts are likely to have occurred. 
We therefore concentrate on the progenitors of present-day UCDs and GCs (i.e. their young counterparts at high 
redshift). 
We construct stellar population models
using the PEGASE\footnote{www.iap.fr/pegase.} \citep{Fioc1997} code to suggest possible
photometric diagnostics to provide observational predictions for upcoming missions such as 
the James Webb Space Telescope (JWST). 
The underlying question for this work is: 
Can a systematic variation of the stellar IMF be confirmed by observations of the likely high-redshift precursors of 
present-day GCs and UCDs using the JWST?

This paper is structured as follows: The first section is devoted to
the introduction of the topic.  Section 2 describes the methods we
use. Section 3 focuses on results and Sections 4 and 5 contain the discussion
and conclusion, respectively.

\section{Methods} 
We compute properties of UCDs and UCD progenitors
assuming that they have formed according to scenario (A),
that is\ by monolithic collapse.  
In such a case we expect that the first UCDs
were formed alongside the  formation of early massive galaxies
($\approx 0.5\,\mathrm{Gyr}$ after the Big Bang).  
Results obtained according to scenario (A) can be used to discuss and constrain
also scenario (B), which is the formation of UCDs through mergers of cluster complexes. 
Since here we assume that GCs are low mass UCDs, we from here on refer to GCs also as (low-mass) UCDs.

\subsection{Parametrization of the UCDs} \label{Sec:red}
For quantifying our approach, we consider UCDs with the following properties: 
the UCD's initial stellar mass, $M_{UCD}\in[10^6
,10^{9}]\,\mathrm{M_{\odot}}$, red-shift, $z\in {0,3,6,9}$,
corresponding, respectively, to ages from the Big-Bang $\approx$ 13.5, 2.1, 0.9, 0.6 Gyr as
demonstrated in Fig.~\ref{fig:tz}.  For the redshift computations
we use the standard $\Lambda$ cold dark matter ($\Lambda$CDM) cosmology with Planck
estimates,
$\Omega_{m}\approx 0.308$, $\Omega_{\Lambda}\approx 0.692$,
$H_0\approx67.8\,\mathrm{kms^{-1}Mpc^{-1}}$
\citep{Planck2016,Planck2016b}.
Other parameters are  metallicity, $\mathrm{[Fe/H]}=-2\,\mathrm{and}\,0$, and the stellar
initial mass function (IMF).

\subsection{The stellar IMF}
We describe the stellar IMF as a
multi-power law, 
\begin{equation}
\xi(m_{\star}) =    \left\{ \begin{array}{ll}
k_1 m^{-\alpha_1} \hspace{1.65cm} 0.08\leq m_{\star}/M_{\odot}<0.50 \,, \\
k_2 m^{-\alpha_2} \hspace{1.65cm} 0.50\leq m_{\star}/M_{\odot}<1.0 \,, \\
k_2 m^{-\alpha_3} \hspace{1.65cm} 1.00\leq m_{\star}/M_{\odot}\leq 120 \,, \\
\end{array} \right.
\label{eq:IMF}
\end{equation}
where 
\begin{eqnarray}
\xi(m_{\star}) = \mathrm{d} N/\mathrm{d} m_{\star}\, 
\end{eqnarray}
is the number of stars per unit of mass and $k_i$ are normalization constants which also ensure continuity of 
the IMF function.
As a benchmark we use the IMF $\alpha$ values derived from the Galactic star forming
regions by \cite{Kroupa2001}, where $\alpha_1=1.3$ and $\alpha_2=\alpha_3 =2.3$, here 
denoted as \textcolor{green}{CAN IMF} (canonical IMF).

A larger $\alpha_1$ than the canonical value leads to a bottom-heavy IMF. 
A bottom-heavy IMF is also described by a single
Salpeter slope $\alpha=\alpha_1=\alpha_2=\alpha_3=2.3$, 
which has been shown to lead to  
slightly elevated $M/L_V$ values \citep{Dabringhausen2008,Mieske2008b} relative to the \textcolor{green}{CAN IMF}. 
Throughout this paper we will call this IMF
the \textcolor{orange}{SAL IMF} (Salpeter IMF).  
An even more bottom-heavy IMF, which might be necessary for
explaining the observed $M/L_V$ values around ten and higher, 
was suggested by \cite{vanDokkum2010} for massive elliptical galaxies, due to features observed in their 
spectra \citep[but see][]{Smith2013}. This IMF is here referred to as the
\textcolor{red}{vDC IMF}, and is characterized by an IMF slope 
$\alpha=\alpha_1=\alpha_2=\alpha_3=3.0$. A dependency of $\alpha$ on star-cluster-scale 
star-formation density and metallicity is currently not known.
\cite{Chabrier2014} suggest increased turbulence to account for a bottom-heavy IMF in dense star forming regions.  However, further theoretical work
\citep{Bertelli2016,Liptai2017} casts doubts on this. This issue clearly needs further
research.

The other option of how to explain the observed $M/L_V$ ratios is a
top-heavy IMF \citep{Dabringhausen2009}.  
The top-heavy IMF has an empirical prescription, which establishes the slope of the heavy-mass end,
$\alpha_3$, over the interval of masses
$(1,120)\,\mathrm{M_{\odot}}$ as a function of metallicity [Fe/H]
and birth density of the embedded star cluster, $\varrho_{cl}$.
The lower stellar masses  ($< 1.0\,\mathrm{M_{\odot}}$) are in our formulation here distributed
according to the \textcolor{green}{CAN IMF}. The relation for $\alpha_3$
by \cite{Marks2012TH} is, 
\begin{equation}\label{eq:MKDP}
\alpha_3=   \left\{ \begin{array}{ll} 2.3 \hspace{2.15cm}
\mathrm{if}\,x<-0.87\,,\\ -0.41x+1.94 \,
\hspace{0.5cm}\mathrm{if}\,x\geq-0.87\,, \end{array} \right.
\end{equation} where \begin{equation} x =
-0.14[Fe/H]+0.99\log_{10}{\left(\frac{\varrho_{cl}}{10^6M_{\odot}pc^{-3}}\right)}\,.
\label{eq:MKrel}
\end{equation} 
These relations have been obtained from a multi-dimensional regression of GC and UCD data. 
To calculate the birth density, $\varrho_{cl}$, we use the empirical
relation from \cite{Marks2012}, where the half-mass radii of embedded
clusters follow 
\begin{equation}
R_{cl}/pc=0.1(M_{ecl}/M_{\odot})^{0.13} \,,\\
\end{equation}
where $M_{ecl}$ is the stellar mass. The total density (gas + star) , $\varrho_{cl}$, 
for a star formation efficiency, $\epsilon$, is given by 
\begin{equation}
  \varrho_{cl} = \frac{3 M_{ecl}}{4 \epsilon \pi R_{cl}^3} \,.
\end{equation}
Further on, we assume a star formation efficiency $\epsilon=0.33$ \citep[e.g.][]{Megeath2016, Banerjee2017} and refer to the 
IMF derived using these relation as the \textcolor{blue}{MKDP IMF}. The $\alpha_3$ 
and half-mass radii $R_{cl}$ are listed in Table~\ref{tab:IMF}. 
Theoretical arguments for the IMF becoming top-heavy with increasing density, temperature, and decreasing metallicity of the star-forming gas cloud have been described by e.g. \cite{Larsen1998}, \cite{Adams1996}, \cite{Dib2007}, \cite{Papadopoulos2010} and \cite{Romano2017}. 

All the IMFs considered here are summarized in Table~\ref{tab:IMFvar}.

\begin{table}[t] \centering \begin{tabular}{c | c | c | c | c |}
IMF & Top-heavy & Canonical & Bott.-heavy  & Bott.-heavy  \\ 
&  \textcolor{blue}{MKDP} &  \textcolor{green}{CAN} &  \textcolor{orange}{SAL} & 
\textcolor{red}{vDC} \\ \hline
$\alpha_1$ & 1.3 & 1.3 & 2.3 & 3.0 \\ \hline
$\alpha_2$ & 2.3 & 2.3 & 2.3 & 3.0 \\  \hline
$\alpha_3$ & 0.6 -- 2.3 & 2.3 & 2.3 & 3.0 \\ \hline
\end{tabular} \centering \caption{Summary of IMF variations used in this paper. The $\alpha$-coefficients 
are defined by Equation~\ref{eq:IMF}. The MKDP IMF depends on the initial stellar mass 
of the system according to Equation~\ref{eq:MKDP}.}
\label{tab:IMFvar}\end{table}

\begin{table*}[t] \centering \begin{tabular}{c c | c cc c |c c} 
& & \multicolumn{4}{c}{MKDP IMF } & \multicolumn{2}{c}{CAN IMF } \\ \hline & & \multicolumn{2}{c}{[Fe/H] = -2} &
\multicolumn{2}{c}{[Fe/H] = 0} & & \\ \hline
$M_{UCD}\,\mathrm{[M_{\odot}]}$ & $R_{cl}\,\mathrm{[pc]}$ & $\alpha_3$
& $10^5N(m_{\star}>8 M_{\odot})$ & $\alpha_3$ & $10^5N(m_{\star}>8
M_{\odot})$ & $\alpha_3$ & $10^5 N(m_{\star}>8 M_{\odot})$ \\ \hline
$10^6$ &    $0.6$ & $1.61$ & $0.2$    &  $1.73$ & $0.2$    & $2.3$ &
$0.1$ \\ \hline $10^7$ &    $0.8$ & $1.37$ & $2.4$    &  $1.48$ &
$2.4$    & 2.3 &  $1.1$ \\ \hline $10^8$ &    $1.1$ & $1.12$ &
$23.2$   &  $1.23$ & $23.8$   & 2.3  &  $10.9$ \\ \hline $10^9$ &
$1.5$ & $0.87$ & $213.8$  &  $0.99$ & $222.7$  & 2.3  &  $109.1$ \\
\hline $10^{10}$ & $2.0$ & $0.62$ & $1943.1$ &  $0.74$ & $2032.6$ &
2.3  &  $1091.2$ \\ \hline 
\end{tabular} \centering
\caption{Values of the initial radii of the embedded star clusters, $R_{cl}$,
\citep{Marks2012}, $\alpha_3$ values (see Eq.~(\ref{eq:MKDP})), and the number of massive stars $N(m_{\star}>8 M_{\odot})$ 
computed as a function of initial stellar mass, $M_{UCD}=M_{ecl}$, and metallicity. For comparison, we also show $N(m_{\star}>8M_{\odot})$ 
for the \textcolor{green}{CAN IMF}. The predicted values of
the embedded cluster/UCD initial radii, $R_{cl}$, are small. However, 
due to stellar and dynamical evolution, UCDs expand by approximately a factor of ten \citep{Dabringhausen2010}. We use the $R_{cl}$ values 
as an empirical extrapolation from GCs and therefore at  UCD scales departures are possible.} 
\label{tab:IMF}\end{table*}

\subsection{PEGASE parameters} 
The above parametrizations are used as input
values to the~P\'EGASE time-dependent stellar population synthesis
code\footnote{www.iap.fr/pegase.} \citep{Fioc1997} and we compute the time
evolution of various quantities, such as the total luminosity, $L_{\rm UCD}$,
number of massive stars ($m_{\star}>8\,\mathrm{M_{\odot}}$), $N_{\rm
mass}$, and also observable properties: the redshifted time evolution of the 
spectral energy distribution (SED),
and colour-magnitude diagrams. Furthermore we investigate in detail the $M/L_V$ values.
Besides the values considered here for the initial stellar mass, metallicity, and redshift of the UCDs
 ($M_{UCD}$, [Fe/H], $z$), we use the following 
values to compute the P\'EGASE models: a conservative value $\eta= 0.05$ for 
the fraction of binaries producing supernovae Ia \citep[e.g.][where $\eta \in (0.05,0.4)$]{Maoz2008}, 
zero inclination, no galactic winds, no in-fall matter, and no metallicity
evolution (this would require a more complicated time-dependent IMF
prescription and a star formation history). 
We take into account nebular emission and we assume no
extinction since it can be corrected based on various assumptions in
the further interpretation of our models. 
It is, however, worth mentioning that
young and intermediate-age massive clusters in 
the local universe seem to be gas- and dust-free very early on 
\citep[e.g.][]{Bastian2014,Longmore2015}.
The star formation history is 
assumed to have the  shape of a step function with a non zero value for the first $5\,\mathrm{Myr}$. 
We will compare this with the instantaneous  star-burst case,
which has the same parameters apart from the fact that all stars form
at the same time. 
\subsection{Calculating $M/L_V$ ratios} 
To evaluate the dynamical mass, $M$, and luminosity in the $V$ band, $L_V$,  several assumptions are made:  
(i) We assume that the UCDs are gas free during the whole period of their evolution, 
which means that our predictions are only valid for pure stellar populations. 
(ii) Mass loss from the UCDs is only through stellar evolution in the form of ejected gas.  
We assume that all stars are kept in the system and no stars are lost by dynamical 
evolution of the UCDs. Stellar loss due to dynamical evolution is an important process \citep[e.g.][]{Balbinot2017} which is, however, 
significant only for systems with initial stellar mass $\lessapprox 10^6\,M_{\odot}$ \citep{Lamers2005,Schulz2016,Brinkmann2017}. 
The reason is that for large stellar mass, $\gtrapprox 10^6\,M_{\odot}$, the tidal radius in a Milky-Way galaxy potential is~$\gtrapprox 100 \, \mathrm{pc}$  for Galactocentric distances~$\gtrapprox\mathrm{few\, kpc}$ such that the vast majority of stars remain bound. This is consistent with using Equation (5) from 
\cite{Lamers2005} which, after integration, gives an upper limit for tidal stellar mass loss over a Hubble time $\approx 10^3\, M_{\odot}$ independent of initial mass. 
(iii) The effect of loss of dark remnants (neutron stars and black holes) on the estimate of the dynamical mass 
for the vDC, SAL, and CAN IMF is negligible to the value of the $M/L_V$ ratio 
(for the CAN IMF dark remnants contribute only few per cent of the UCD mass). 
However, in the MKDP IMF case, where
dark remnants contribute a substantial fraction to the total mass of the system, the amount of dark remnants kept 
is no longer negligible. 
The actual fraction of dark remnants kept in a system is still being discussed mainly because to study remnant ejections,  
close dynamical encounters in a system are important on a Hubble timescale \citep{Banerjee2017}. Such studies are extremely 
computationally intensive for systems as massive as UCDs. On the other hand, \cite{Peuten2016} and
\cite{Baumgardt2017}, constrain the retention fraction of dark remnants in lower mass GCs,
which allows us to deduce implications for UCDs.

\subsubsection{Retention fraction of dark remnants}
The mass-to-light ratios of UCDs depend on the retention
fraction of stellar remnants within the system once the
progenitor stars die. In order to assess the possible range
of retention fractions, we assume that white dwarfs (WDs)
receive no kicks upon the death of their progenitor stars
such that all WDs remain bound to  the system. A star more
massive than $8\,M_\odot$ explodes as a type II supernova or
implodes leaving either a neutron star or a stellar black
hole. The kicks these receive during such violent events due
to the asymmetry of the explosion or implosion are uncertain.
Large natal kicks will lead to the loss of most such
remnants from the system. 

We estimate the retention fraction by assuming that 10~per~cent 
of all neutron stars and black holes are retained in a
globular cluster with a mass of $10^5\,M_\odot$. This is a
conservative assumption as \cite{Baumgardt2017} and 
\cite{Peuten2016}  constrain the retention fraction to be less than about
50~per cent for globular clusters (GCs)  based on a detailed
study of their observed mass segregation. With this
normalisation condition, and assuming two possible
radius-mass relations for GCs and for UCDs, we can estimate
the likely values of the retention fraction of stellar
remnants (neutron stars and black holes) as a function of
birth system mass, $M$. One possibility for the $R(M)$
relation is to assume the observed radii of clusters
(typically about $3\,$pc) and of UCDs 
\citep[Eq. (4) in][]{Dabringhausen2008}. 
The other possibility is to assume that the
stars die in their birth systems before these expand \citep{Dabringhausen2010} 
due to residual gas expulsion and stellar-evolution-driven mass
loss with the $R(M)$ relation constrained by Marks \& Kroupa
(2012, their Eq.~7).  Given the $R(M)$ relation, we assume that remnants are
lost if their speed after the kick is larger than the
central escape velocity. We assume the systems to be Plummer
models and that the kick velocities follow a
Maxwell-Boltzmann distribution (MBD) with velocity
dispersion $\sigma_{\rm kick}$. 

\begin{figure}[ht!] \begin{center}
                \scalebox{1.0}{\includegraphics{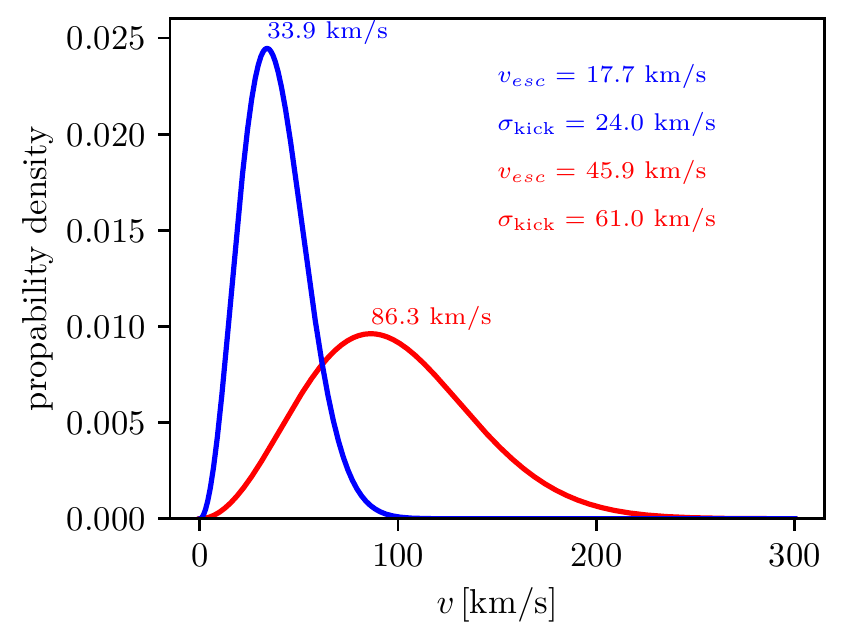}}
        \end{center}
        \caption{
                Distribution function of speeds of dark stellar remnants.
                We use the conservative estimate that only 10 per cent of dark
                remnants are kept in $10^5\,M_{\odot}$ system due to  
                SN kicks. We constrain the SN kick velocity distribution
                assuming it has a Maxwell-Boltzman shape for birth radius 
                \citep[][red curve]{Marks2012} and for present day radius, $3$ pc for
                GCs and for the UCDs we use the mass-radius relation from
                \cite{Dabringhausen2008} (their Eq. 4, blue curve). The peak velocity, $\sigma_{\mathrm{kick}}$ , and 
                $v_{esc}$ values are stated.
        } 
        \label{fig:dispersion} \end{figure}

\begin{figure}[ht!] \begin{center}
                \scalebox{1.0}{\includegraphics{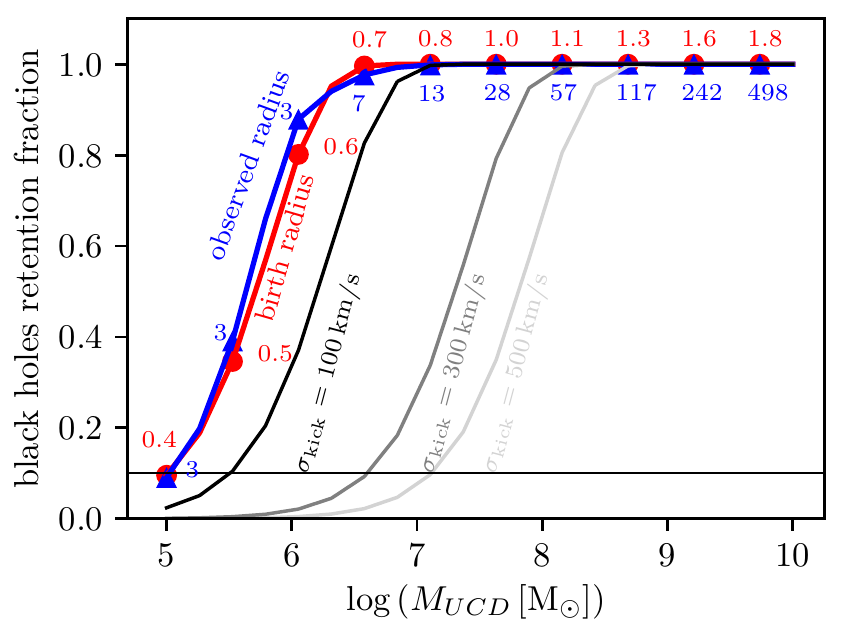}}
        \end{center}
        \caption{Retention fraction
                of dark remnants as a function of system mass if we assume that the
                main mechanism of dark remnants ejection are 
                SN kicks with a Maxwell-Boltzman kick velocity distribution shown in Fig.~\ref{fig:dispersion}. 
                We assumed two different mass-radius relations (red and blue curves, as in Fig. \ref{fig:dispersion}), radii for each mass being written 
                next to the corresponding points and a normalization described in the text. In black and grey we plot retention fractions if larger 
                kick velocity dispersions are assumed. In this case we use the mass-radius relation from \cite{Marks2012}.
        } 
        \label{fig:retention} \end{figure}

Figure~\ref{fig:dispersion} shows the resulting MBD for the two assumed $R(M)$
relations. The adopted normalization condition forces the
MBD to be narrow, with a small kick, $\sigma_{\rm kick}= 24 \,$km/s,
for the present-day radii of clusters and UCDs, while
applying the birth radii results in a larger kick, $\sigma_{\rm
        kick}= 61\,$km/s. Despite the two different values,
the resulting retention fractions, defined for the present
purpose as the fraction of remnants with a speed smaller
than the central escape speed, are very similar. This is
shown in Fig.~\ref{fig:retention}, from which it also follows that the retention
fraction increases steeply with system mass such that for
$M>10^6\,M_\odot$ the retention fraction can be assumed to
be near to 100~per cent, if the normalization condition
applied here holds. We also plot the retention fraction for 
distributions with larger $\sigma_{\mathrm{kick}}$ values, namely $100$, $300,$ and
$500$ km/s. Given these results it is reasonable to assume that the retention 
fraction of UCDs with birth masses larger than $10^7\,M_{\odot}$
is close to 100\%. 
Indirect evidence suggesting a possible large retention fraction in UCDs are recent findings of super massive 
black holes (SMBHs) in UCDs \citep[e.g.][see Sec. \ref{Sec:MLv}]{Seth2014, Ahn2017}.
Nevertheless, we still investigate the effect of smaller retention fractions on the $M/L_V$ values 
to allow us to see the maximum impact the assumed IMF can potentially have.

\subsection{Limitations of the models}
We note here three possible limitations to the models introduced above:
\begin{enumerate}
        \item \label{sec:binevol} Effects of binaries\\
        The majority of stars form in binaries
         \citep[e.g.][]{Marks2011, Thies2015}.
         Mass transfer and mergers can rejuvenate stellar populations leading to significantly more UV radiation even after a few Gyr \citep[][BPASS code]{Stanway2016}. 
         Since here we focus on the IR region
         mostly and on systems younger than 100 Myr, it is likely that our conclusions are not affected strongly if binary-star evolution is taken into account, but it is nevertheless important to check this factor in the future.
    
  \item Multiple populations\\
  It is well established that GCs have multiple stellar populations
        \citep[e.g][review]{Renzini2015}. This may be true for UCDs as well.  Future SED modelling has the potential
        to address how and if these can be observed in young systems at high
        redshift. However, we caution that binary
        stellar evolution (Item~\ref{sec:binevol} above) may lead to degeneracies.

\item Statistical importance\\
    Large
        statistical samples may be needed to ascertain a systematic variation of the IMF with physical conditions  (\cite{Dabringhausen2008,Dabringhausen2009,Dabringhausen2012}).
\end{enumerate}

\section{Results}

 For the grid of
chosen parameters we construct a time grid (1-10 Myr with 1
Myr step, 10-100 Myr with 10 Myr steps, 100-1000 Myr with 100 Myr
steps, and 1-13 Gyr with 1 Gyr steps) of SEDs, which contain
all the light information from the source necessary for the
construction of other observables.
 Further details (e.g comparison of the PEGASE code and the StarBurst99--code \citep{Leitherer1999} with description 
of the evolution of SEDs with time and dependency on redshift) can be found in Appendix \ref{app:extSED}.

\subsection{Evolution of the bolometric luminosity with time}
\begin{figure*}[ht!] \begin{center}
\scalebox{1.0}{\includegraphics{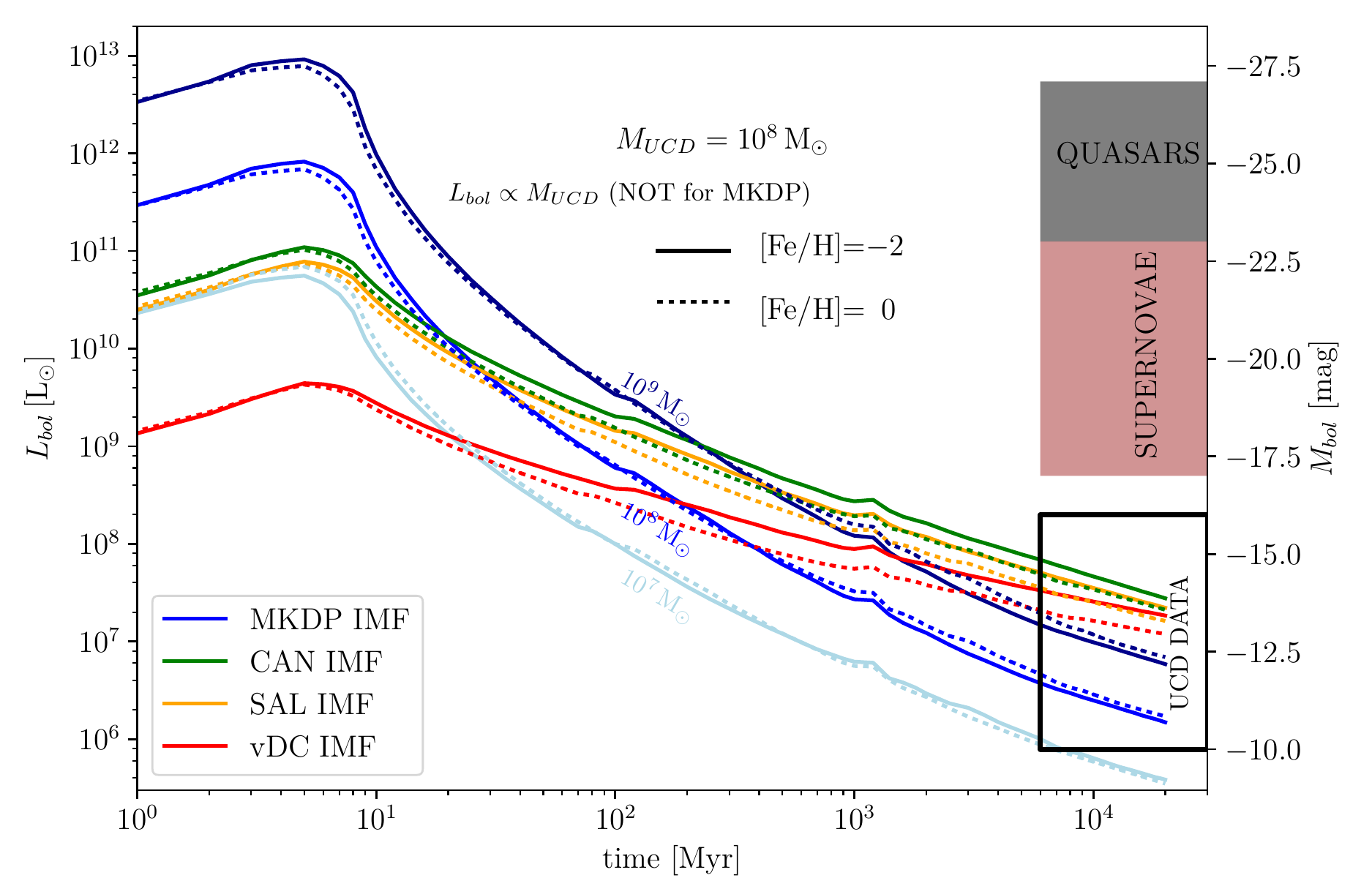}} \end{center}
\caption{Time evolution of the bolometric luminosity for different IMFs. 
The MKDP IMF changes with the initial mass of the UCD and
does not scale linearly with its mass. In contrast, the vDC, SAL, and CAN IMFs 
have UCD-mass-independent slopes and scale proportionally with $M_{UCD}$. The grey 
panel shows the typical luminosities of quasars \citep{Dunlop1993,Dunlop2003,Souchay2015} and 
the brown panel shows the luminosity span for the peak luminosities supernovae \citep{Gal-Yam2012, Lyman2016}, which might cause 
luminosity variations in $L_{bol}$ of UCDs younger than about 50 Myr according to SNe rates.
} \label{fig:Lbol} \end{figure*}

To demonstrate how luminous progenitors of UCDs could have been, 
we compute the bolometric luminosity, $L_{bol}$, as a function of time. 
The $L_{bol}(t)$ dependency is shown in~Fig.~\ref{fig:Lbol}. 
The benchmark initial mass is $M_{UCD}=10^8\,M_{\odot}$. 
For the cases of the \textcolor{green}{CAN}, \textcolor{orange}{SAL,} and 
\textcolor{red}{vDC} IMF, the $L_{bol}$ values are proportional to $M_{UCD}$. The \textcolor{blue}{MKDP}
IMF is a function of $M_{UCD}$ and therefore we plot $L_{bol}(t)$ also for 
$M_{UCD}=10^7\, \mathrm{and}\, 10^9\,M_{\odot}$. The main difference between different \textcolor{blue}{MKDP} IMFs 
(for different $M_{UCD}$) is the slope of the luminosity as a function of time. 
For a more top-heavy IMF, the decrease of the luminosity with time is steeper. 

There are several noticeable features in this figure: (i) for the \textcolor{blue}{MKDP} IMF, UCDs with initial 
stellar mass $M_{UCD} \gtrapprox 10^8\,M_{\odot}$ are as bright as quasars 
\citep{Dunlop1993,Dunlop2003,Souchay2015} for the first few $10^7$ yr. We use low-redshift quasars because 
for these we were able to find bolometric luminosities, but also absolute magnitudes in the $V$ and $I$ band. 
We note, however, that high redshift quasars show very comparable luminosities as shown for example by \cite{Mortlock2011}.
This is mainly due to the presence of a large number of O and B stars ($10^6-10^7$) in the  system. 
Due to stellar evolution and core-collapse supernova explosions, these UCDs will be variable on a timescale of months. As already suggested by \cite{Terlevich1993}, such objects might be confused with 
quasars, especially for example in large photometric surveys. 
(ii)  After less than a $100$ Myr, a strong degeneracy between the  IMF and $M_{UCD}$ appears. That is, for 
different $M_{UCD}$ and different IMFs, similar luminosities comparable to those of observed UCDs occur. 
(iii) The metallicity is a second-order effect.

\begin{table}[t] \centering \begin{tabular}{c | c | c | c | c }
QSO & $z$ & $M_V$ & $M_I$ & RQ/RL \\ \hline
0923+201& 0.193 & -24.6 & -24.28 & RQ \\ \hline
2344+184& 0.138 & -23.6 & -23.80 & RQ \\ \hline
1635+119& 0.147 & -23.1 & -22.39 & RQ \\ \hline
1012+008& 0.185 & -24.3 & -25.39 & RQ \\ \hline
1004+130& 0.241 & -25.7 & -24.90 & RL \\ \hline
1020-103& 0.197 & -24.2 & -24.35 & RL  \\ \hline
2355-082& 0.211 & -23.0 & -23.71 & RL \\ \hline
\end{tabular} \centering \caption{Quasar (QSO) data. The 
  first column is the QSO identification, $z$ is the redshift, $M_V$ and $M_I$ are 
  absolute magnitudes, and RQ/RL labels the QSO as being radio quiet (RQ) or radio 
  loud (RL). 
We have not found any publication or catalogue presenting both the 
absolute magnitudes in the V filter, $M_V$, and I filter, $M_I$. Therefore, the data 
in the table use the catalogue by \cite{Souchay2015} containing the 
$M_I$ values. The $M_V$ values are taken from \cite{Dunlop1993} and 
\cite{Dunlop2003}.
}
\label{tab:QSO}\end{table}

\subsection{Time evolution of the $M/L_V$ ratio} 
\begin{figure*}[ht!] \begin{center}
\scalebox{1.0}{\includegraphics{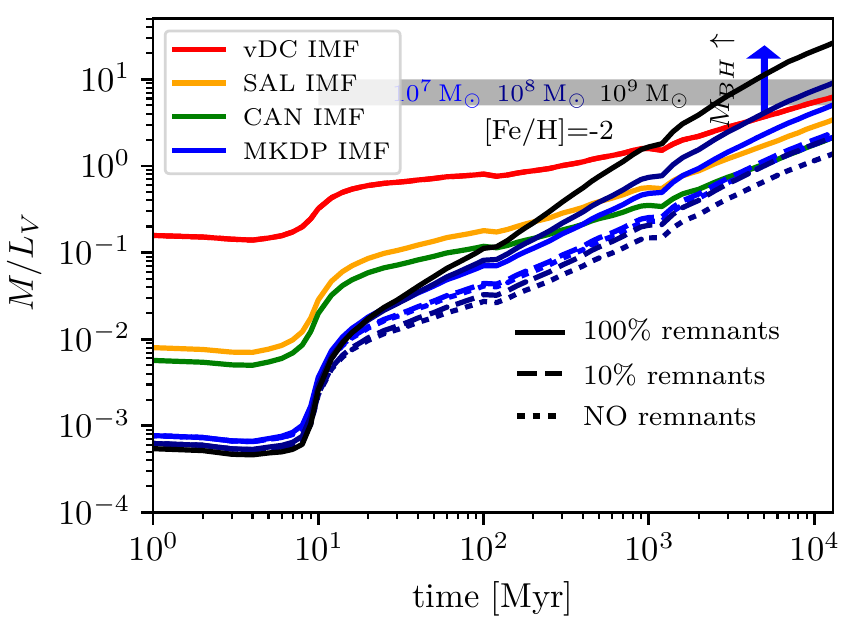}}
\scalebox{1.0}{\includegraphics{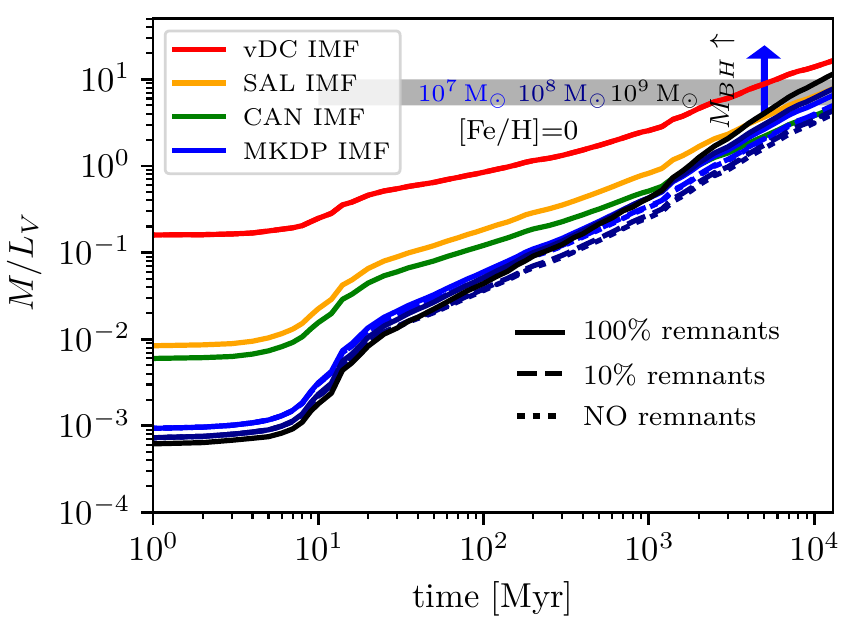}} \end{center}
\caption{Time evolution of $M/L_V$ for all IMFs considered (vDC,
SAL, CAN, MKDP).  We assume that the system is gas free and that neither stars
nor remnants are leaving (the dashed lines show the change of the $M/L_V$ value if only 
certain fractions of remnants are kept). The upwards-pointing blue arrow demonstrates that if we would assume 
more massive remnants (e.g. due to implosion directly to a BH without a SN explosion), 
then this would lead to larger values of $M/L_V$.
The models are computed for the
case of $M_{UCD}=10^9\,\mathrm{M_{\odot}}$, however, results are mass independent. 
The only exception is the MKDP IMF, which depends on  initial object mass, $M_{UCD}$, and also
contains a large fraction of mass in high mass stars, and is plotted for
$M_{UCD}=$ ($10^9,\,10^8,\,10^7\,\mathrm{M_{\odot}}$, bottom to top) and different fractions of
remnants retained (100\%,10\%,0\%).  {\bf Left panel:} The
evolution for [Fe/H]=-2. {\bf Right panel:} The evolution for
[Fe/H]=0. The grey band indicates the span of the observed
present-day $M/L_V$ values, approximately five to ten for the majority of UCDs 
\citep[][]{Mieske2008, Dabringhausen2009}.  The scales are identical for both panels.} \label{fig:MtoLV} \end{figure*}

The PEGASE output allows us to
evaluate the mass-to-light ()$M/L_V$) ratios in arbitrary photometric filters since
with the time evolution the code keeps information about the current
stellar mass, mass in black holes and neutron stars, and also about the
gas (non-consumed initial gas and the gas ejected by stars).  Since
the vDC or SAL (bottom-heavy) IMFs do not depend on the initial UCD mass, 
neither does the $M/L_V$ ratio. The
situation is different for the MKDP IMF and we therefore plot the time
evolution of the $M/L_V$ ratio for different initial masses in Fig.~\ref{fig:MtoLV}.

\subsubsection{$M/L_V$ ratio variations}
The results show that the largest differences in
$M/L_V$ ratios are evident in the first $\approx 100\,\mathrm{Myr}$, which is the
age corresponding to the time when the most massive stars evolve into
dark remnants (see Fig.~\ref{fig:MtoLV} where we can also
see the effect of the retention fraction of remnants on the
$M/L_V$ values). For later times (> 100 Myr), degeneracy among the data appears: 
the UCDs formed with a MKDP IMF are equally or less bright
than the UCDs with a CAN or SAL IMF
formed with the same initial mass. Even the $M/L_V$ ratios might become indistinguishable 
at times older than $\approx 100\,\mathrm{Myr}$ for different IMFs within the observational 
and metallicity uncertainties. 

\begin{figure*}[ht!] \begin{center}
\scalebox{1.0}{\includegraphics{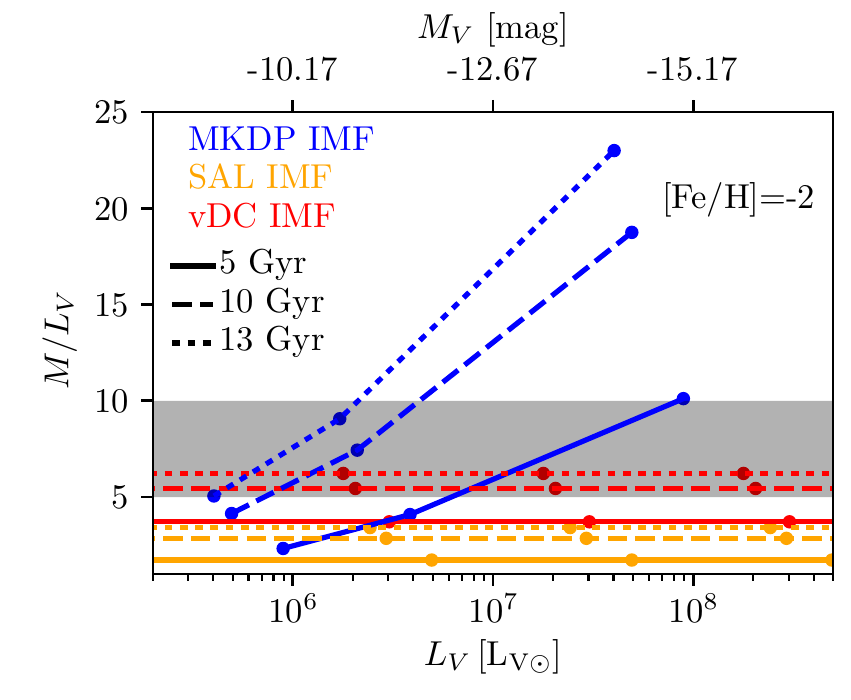}}
\scalebox{1.0}{\includegraphics{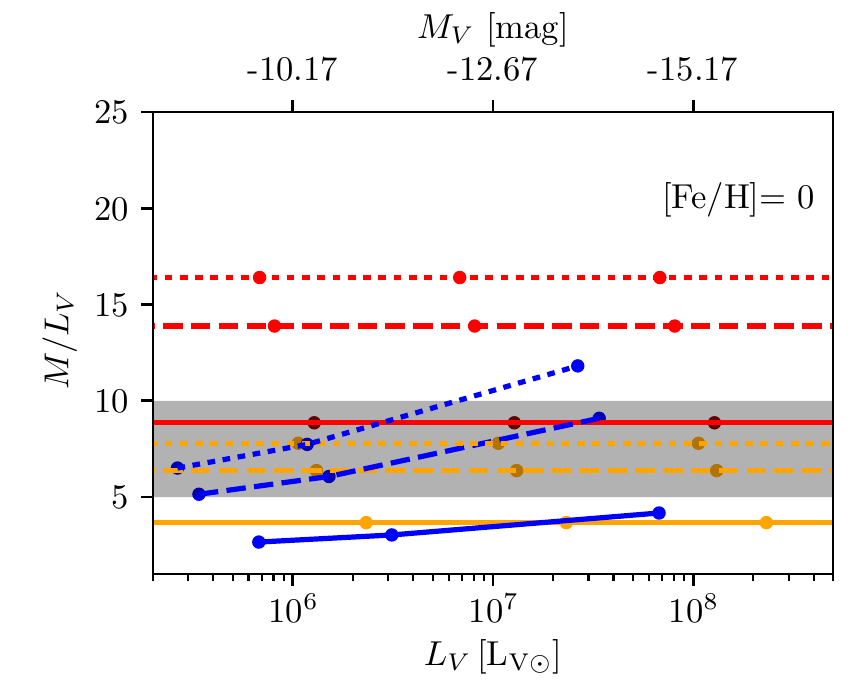}} \end{center}
\caption{Values of the $M/L_V$ ratio as a function of $L_V$ for three different times
$5,\,10,\,13\,\mathrm{Gyr}$.  The case of the MKDP IMF is not constant
since the MKDP IMF is a function of initial mass. The plotted curves assume
all remnants are kept; if only a fraction are kept, the $M/L_V$ values would
become smaller accordingly.  There are two horizontal axes, the one on
the top of the plots shows corresponding values of the absolute magnitude in the rest-frame 
V band. The observed values of $M/L_V$ for UCDs 
are in the interval from approximately five to ten, with few 
values spanning up to 15 \citep{Mieske2002}, which are shown by the grey band. 
The points plotted on the curves mark luminosities and $M/L_V$ values for UCDs starting with 
initial mass of $10^9, 10^8$ , and $10^7 \, M_{\odot}$. 
 {\bf Left panel:} [Fe/H]=-2.{\bf \ Right
panel:} [Fe/H]=0.} \label{fig:MtoLVLV} \end{figure*}

Using the same models, we construct the dependence of $M/L_V$ on
$L_V$ for different evolutionary times (Fig.~\ref{fig:MtoLVLV}). 
To construct these plots, we assumed the set of
initial UCD masses to be $10^7,\, 10^8,\, 10^9\,\mathrm{M_{\odot}}$.
As is clear from the comparison of the panels in Fig.~\ref{fig:MtoLVLV},
the metallicity has a large effect on the $M/L_V$
values. 

\subsection{The supernova rates} 
\begin{figure}[h] \begin{center}
\scalebox{1.0}{\includegraphics{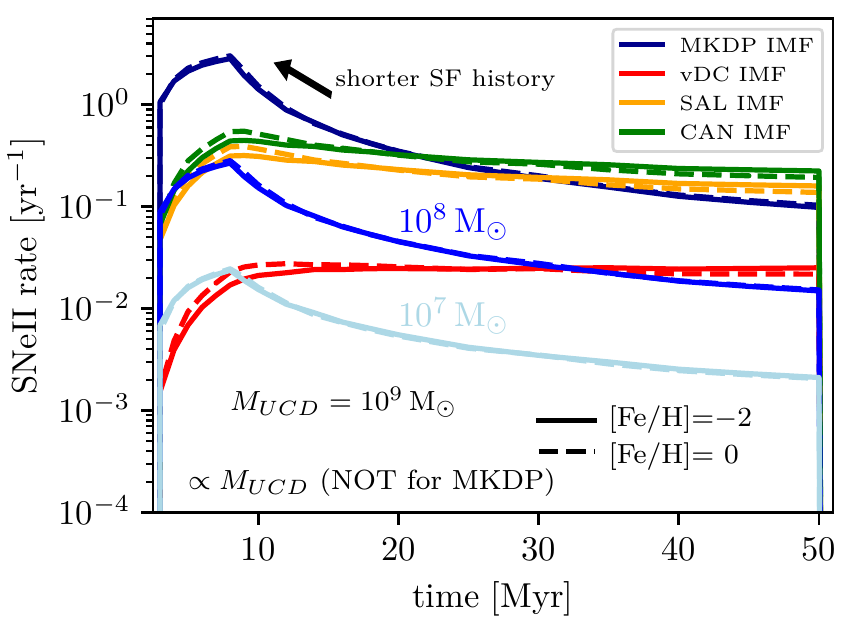}}
\end{center}
\caption{SNe II rate as a function of time for 
a constant star formation history over a period of 5 Myr. In the instantaneous star-burst 
case, the SNeII rate may be$10\times$ higher at ages $<10 \,\mathrm{Myr}$.
We show the case of the vDC, SAL, CAN, and MKDP IMFs. Since only the MKDP IMF does not scale linearly with initial stellar mass, 
$M_{UCD}$, we show also the lines for $M_{UCD}= 10^7\,\mathrm{M_{\odot}}$ and
$10^8\,\mathrm{M_{\odot}}$; for the other IMFs we chose to plot only
$10^9\,\mathrm{M_{\odot}}$. The arrow indicates that the peak of the SN II rate shifts to the left and upwards for a star formation history which is shorter than 5 Myr. } \label{fig:SNIIrate} \end{figure}

The supernovae (SNe) II rate
\citep{Lonsdale2006,Anderson2011} can be a very good indicator of the
IMF as one can see in Fig.~\ref{fig:SNIIrate}. According to the standard stellar evolutionary
tracks employed here, every star more massive than
$8\,\mathrm{M_{\odot}}$ ends as a  supernova explosion. 
However, this may  not always be 
the case. As the metallicity varies it may happen that a star of a given mass may 
implode and create a black hole directly without any explosion
\citep[see e.g.][]{Pejcha2015} and therefore our theoretical prediction
represents the upper limit to the SN II rate. 

It is important to point out that the SN rate depends on the star formation history of 
a system. If the whole system is formed during an instantaneous starburst, the peak SN II rate 
might by a factor of ten higher than in the case of constant star formation over a period of 5 Myr. 
On the other hand, if the star formation is more extended then the period of high SN II rate lasts longer. 

Considering the luminosities of SNe \citep{Gal-Yam2012,Lyman2016}, in Fig.~\ref{fig:Lbol} we show that at the later phases, 
$> 10$ Myr, the SNe can reach up to 10 per cent of the UCD's  total flux
for the \textcolor{blue}{MKDP} IMF and $10^9\,M_{\odot}$ initial mass (for smaller initial masses, SN II explosions will be more pronounced)
and therefore might be detectable as photometric fluctuations on the scale of months.

\begin{figure}[h] \begin{center}
                \scalebox{1.0}{\includegraphics{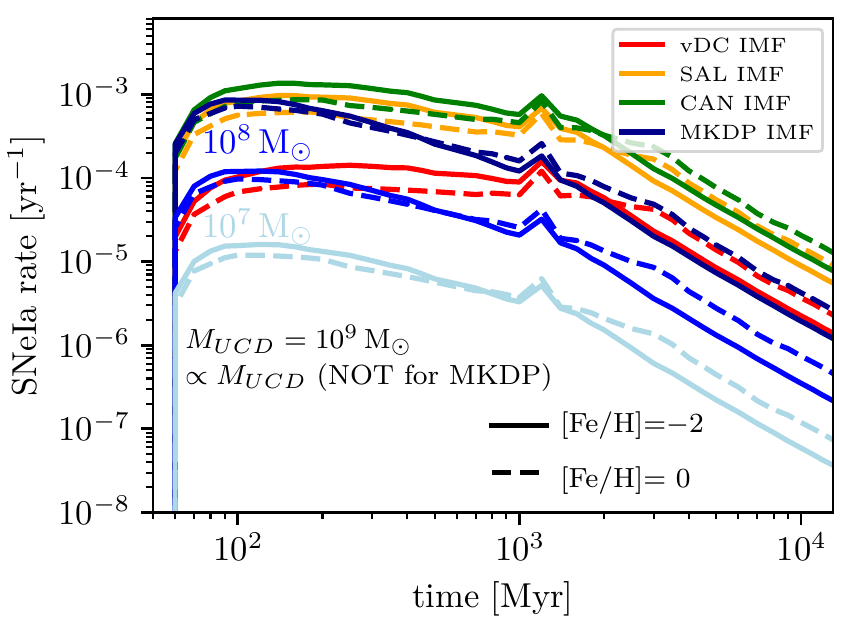}} \end{center}
        \caption{SNe Ia rate as a function of time. 
                We show the case of the vDC, SAL, 
                CAN, and MKDP IMFs. Since only the MKDP IMF does not scale linearly with initial stellar mass, 
                $M_{UCD}$, we show also the lines for $M_{UCD}= 10^7\,\mathrm{M_{\odot}}$ and
                $10^8\,\mathrm{M_{\odot}}$; for the other IMFs we chose to plot only
                $10^9\,\mathrm{M_{\odot}}$.  The text gives more details. } \label{fig:SNIarate} \end{figure}

To compute the SN Ia, rate we adopt a
conservative fraction, $\eta = 0.05$, of intermediate-mass stars that 
eventually explode as SNe Ia. According to \cite{Maoz2008},
$\eta = 0.02 - 0.4$.  We use $\eta = 0.05$ also because it is a default 
value of this parameter and therefore our results will probably be comparable 
with other studies. The $\eta$ value affects
the estimates of the number of SNe Ia and also of the ejecta
\citep{Thielemann1986,Greggio1983,Matteucci1986}. The computed rates 
are plotted in Fig.~\ref{fig:SNIarate}. 
Using a value as large as  $\eta = 0.4$ \citep[][upper value]{Maoz2008} 
would imply an eight times larger SN Ia rate. The other results, such as the luminosities or $M/L_V$ values, are not affected by the value of $\eta$, since different values of $\eta$ do not change the mass distribution of stars.

\subsection{Time evolution of the $\beta_{UV}$ slope} \label{sec:betaUV}
\begin{figure*}[ht!] \begin{center}
                \scalebox{1.0}{\includegraphics{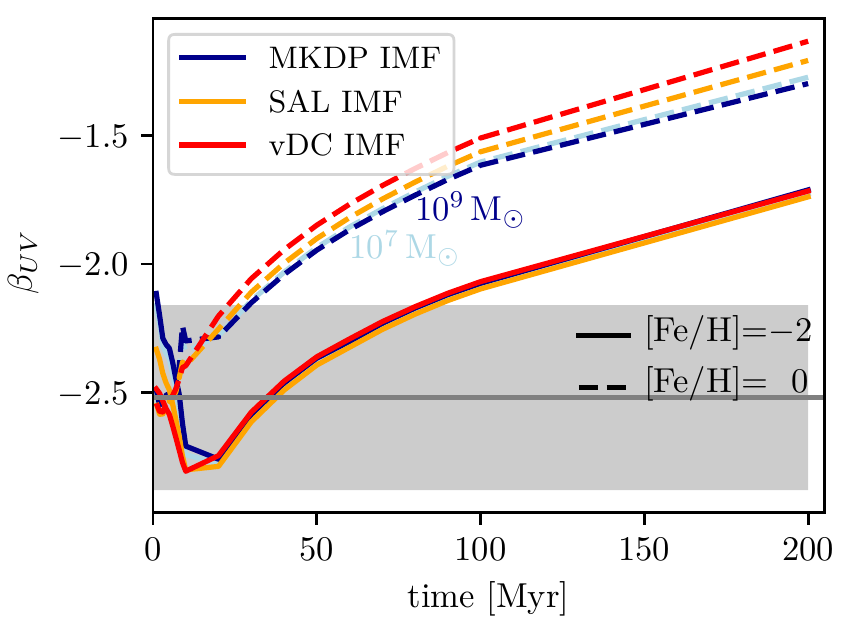}}
                \scalebox{1.0}{\includegraphics{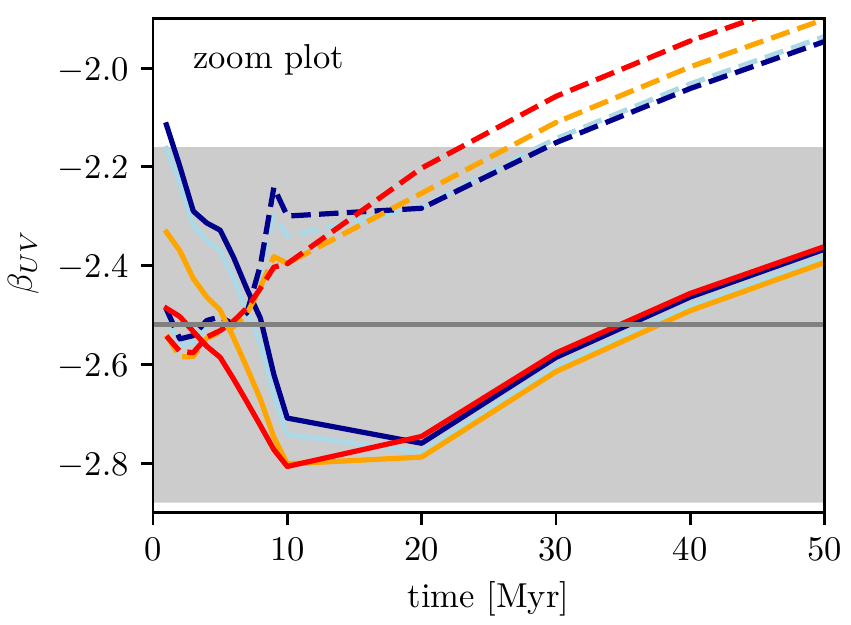}} \end{center}
        \caption{Time evolution of the $\beta_{UV}$ values fitted over the
                wavelength interval (1350,3500)\AA$ $ to the SEDs in the rest frame
                of the observed object in units
                $\mathrm{ergs^{-1}cm^{-2}\AA^{-1}}$. The $\beta_{UV}$ slope is
                $M_{UCD}$ -independent, but not for the case of the MKDP IMF for which
                we plot the case of $M_{UCD}=10^7\,\mathrm{M_{\odot}}$ and
                $10^9\,\mathrm{M_{\odot}}$. {\bf Left panel:} The time evolution of
                the $\beta_{UV}$ slope over the time period of 200 Myr. {\bf Right
                        panel:} The zoom-in plot covers the first 50 Myr. The grey band
                shows the measurement from \cite{Vanzella2017} for their object
                GC1. } \label{fig:betaUV} \end{figure*}

The $\beta_{UV}$
slope is defined here as the slope of the fitted linear function to the
logarithmically scaled SED, expressed in units of
$\mathrm{ergs^{-1}cm^{-2}\AA^{-1}}$, in a wavelength interval
(1350,3500)\AA$ $ in the rest frame of the observed object, as
shown in Fig.~\ref{fig:beta_fit} in the Appendix.
We computed $\beta_{UV}$ for the complete set of our SEDs
(Fig.~\ref{fig:betaUV}).  The $\beta_{UV}$ values have been determined
for objects down to $10^6\,M_{\odot}$ (for a top-heavy IMF this may be $10^5\,M_{\odot}$) at high redshifts (up to $z=6$)
\citep{Vanzella2017} and therefore might allow very useful additional
constraints on the IMF and the age estimates. The $\beta_{UV}$ values
are metallicity sensitive and have a generally increasing trend
with age. The dependence on the IMF is stronger for objects younger
than 10 Myr and at low metallicity, $\beta_{UV}^{MKDP} \approx -2.1$,
$\beta_{UV}^{vDC} \approx -2.6$ at age 1 Myr for [Fe/H]=-2.  At an age
of 200~Myr, these values evolve to $\beta_{UV}^{MKDP} \approx -1.3$,
$\beta_{UV}^{vDC} \approx -1.1$.

At a high redshift it may not always be possible to obtain a spectrum
of a UCD. In such a case we can obtain photometric fluxes in at least
two filters and use these to approximate the $\beta_{UV}$ slope. For
$z=3$ one could, for example, use the JWST NIRCam instrument with
filters F070W and F115W. For $z=6$ and $z=9$ it is possible to use the
same instrument but with filters F090W, F200W, and F115W, F300M,
respectively.

\subsection{The colour-magnitude and the colour-colour diagrams} 
\begin{figure*}[ht!] \begin{center}
                \scalebox{1.0}{\includegraphics{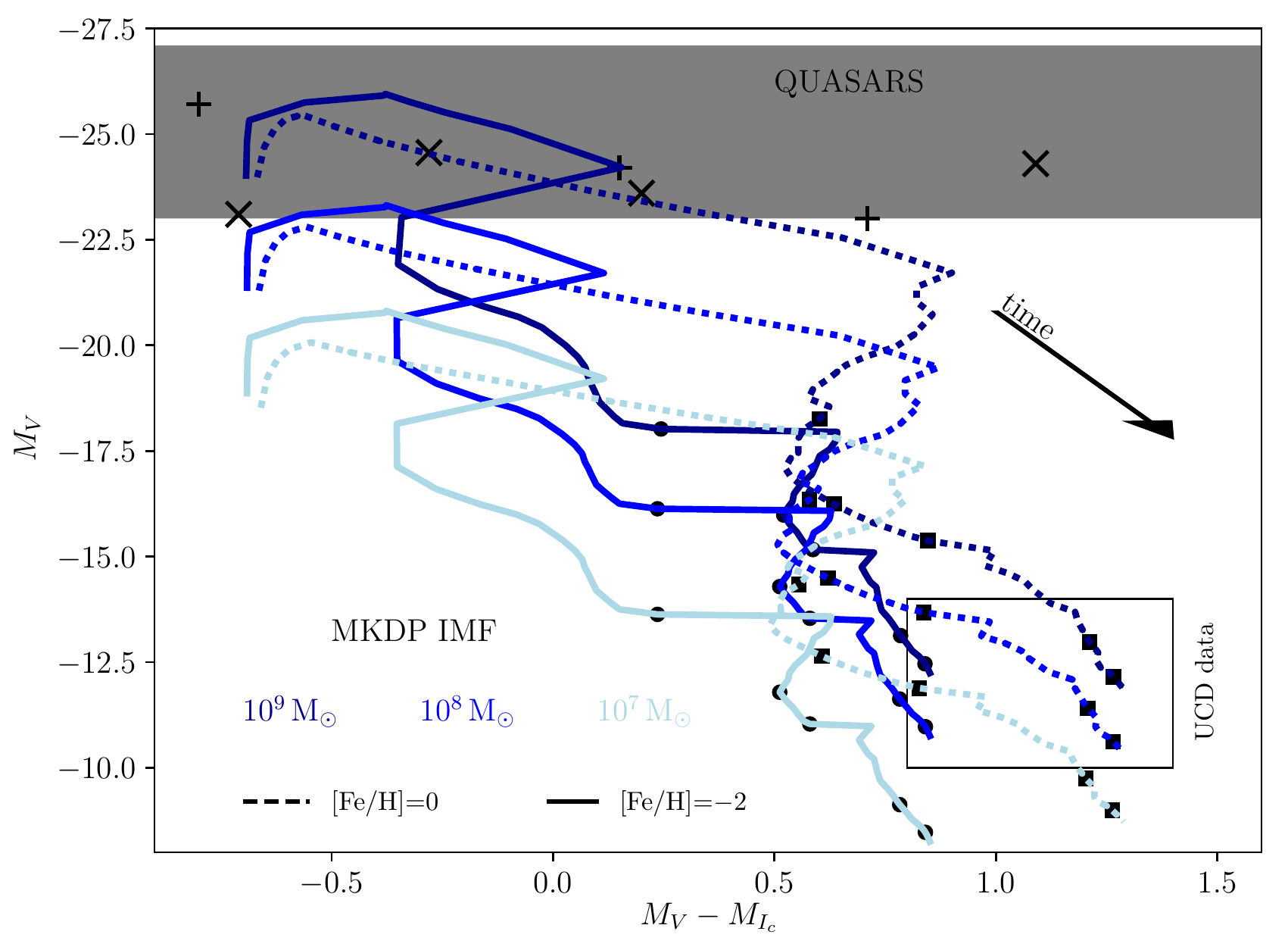}}
        \end{center} \caption{Colour-magnitude diagram showing $M_V$ as a
                function of $V-I_c$ (in the UCD rest frame). These latter are photometric
                filters directly computed by PEGASE. We consider
                $10^7,\,10^8,\,10^9\,\mathrm{M_{\odot}}$ as initial stellar masses,
                and metallicity values $[Fe/H]=-2,\,0$.  Here we show only the
                results for the MKDP IMF plotted together with the quasar data (black
                cross and plus markers, cross for radio quiet and plus for radio
                loud quasars) from \cite{Dunlop1993,Dunlop2003,Souchay2015}. The
                data are compiled in Table~\ref{tab:QSO}. The CAN IMF, SAL IMF and
                vDC IMF are shown in Fig.~\ref{fig:McolD}. The arrow indicates the
                time evolution for the UCD models; the black filled circles and
                squares mark evolutionary time, from left to right: 100 Myr, 500
                Myr, 1 Gyr, 5 Gyr, 10 Gyr for [Fe/H]=-2 and 0 dex. The rectangular
                region indicates where the majority of observed UCDs are located
                \citep[e.g.][]{Evstigneeva2008}.  Since UCDs have a different
                metallicity, we do not plot individual data points.
        } \label{fig:McolD_MKDP} \end{figure*}

\begin{figure}[ht!] \begin{center}
                \scalebox{1.0}{\includegraphics{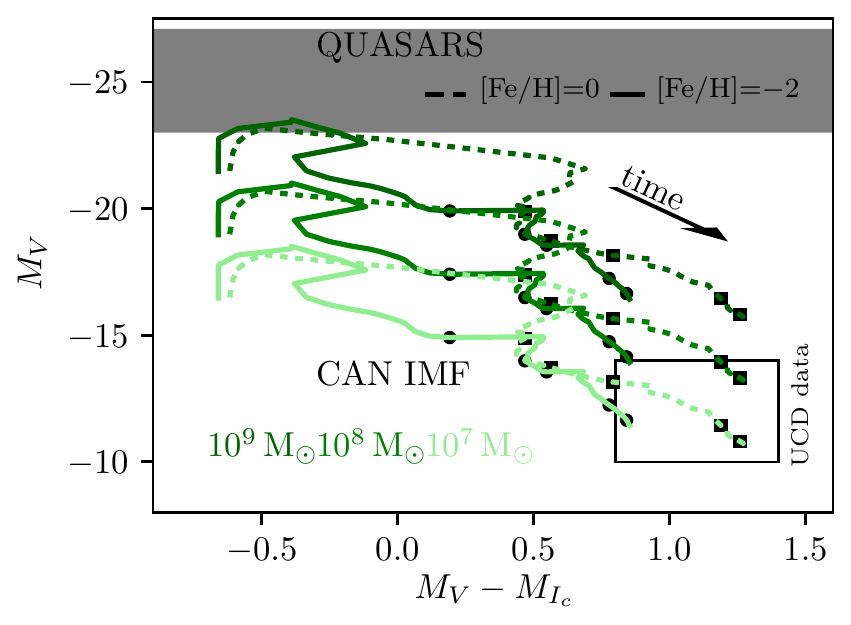}}
                \scalebox{1.0}{\includegraphics{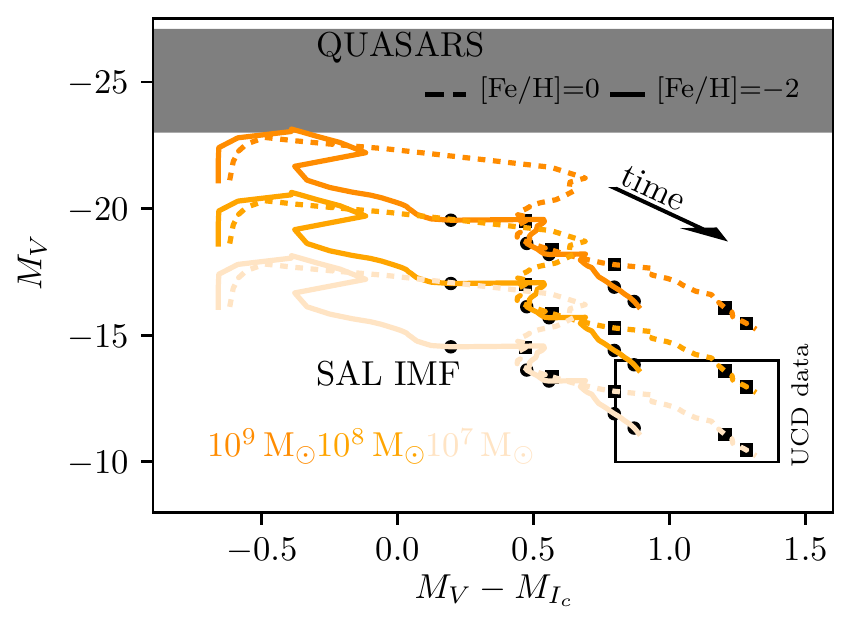}}
                \scalebox{1.0}{\includegraphics{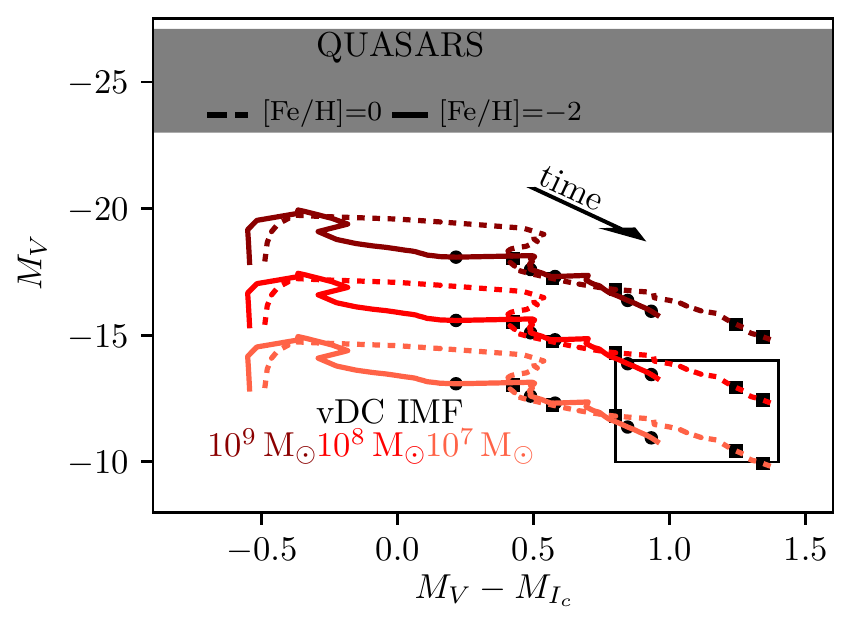}}
        \end{center} \caption{As Fig.~\ref{fig:McolD_MKDP} but for the CAN IMF (MIRI {\bf Top panel)}, the SAL IMF ({\bf middle panel),} and the vDC IMF {\bf
                        (bottom panel)}.} \label{fig:McolD} \end{figure}

Other observables, which can be computed from the SED and for standard
filters and which are provided by PEGASE, are various colours and magnitudes.
The time evolution of our objects is shown in the $V$ versus $V-I_c$
diagram for comparison with other work
\citep[e.g.][]{Evstigneeva2008}.  This is done for the CAN, SAL, and
vDC IMFs in Fig.~\ref{fig:McolD}, and for the MKDP IMF in
Fig.~\ref{fig:McolD_MKDP}. As expected we can see the increasingly
strong degeneracy with time that makes it hard to distinguish the
metallicity, initial mass, or the IMF.

The colour-colour diagram has the advantage of having differences in
magnitudes on both axis and therefore in a way subdues the information
about the absolute values and enhances the differences in spectral
shapes and features. As is shown in Fig.~\ref{fig:col_col}, we chose to
use the standard filters J, K, and N. The N filter is almost identical
to F1000W on the mid-infrared instrument (MIRI) on the JWST. The J and K filters
cover a similar range as the F115W and F200W filters on near infrared camera (NIRCam) on the
JWST. The results presented here assume transmission functions
corresponding to their filters to be given by the rectangular regions
shown in Fig.~\ref{fig:SED_RED}.  These filters possess several useful
characteristics: (i) even for a redshift of value nine, they are still in
the spectral range covered by the PEGASE SEDs and therefore we do not
introduce any additional errors by using the extrapolated range of the
SEDs, (ii) J, K, and N cover large parts of the SEDs and therefore are
good representatives of an overall shape, and (iii) the N filter is
still in the well-described region where we do not expect large
discrepancies from model to model and therefore also in real
measurements as can be seen in Fig.~\ref{fig:SED_SB99}.

\subsection{Predictions for JWST}
\label{sec:JWSTpreds}
The first question which we need to ask is whether UCD progenitors are
bright enough to be detected and if so up to which redshift. To
quantify this we use the upcoming James Webb Space Telescope (JWST) as
a benchmark. To cover the wavelength region computed here by PEGASE,
the most suitable instrument is NIRCam in imaging mode, in total
covering the region from 0.6 to 5 $\mathrm{\mu m}$. To probe
longer wavelengths, we also compute predictions for the MIRI instrument
in imaging mode.

All predictions we make are for the UCDs with an initial stellar mass
of $10^8\,M_{\odot}$; we use the CAN IMF as the standard and consider
also the MKDP IMF.  For the NIRCam instrument we use as a setup sub
arrays FULL, readout DEEP8, groups 10, integration 1, and exposures
5. This results in a total exposure time of 10149 s. For the MIRI
instrument the parameters are: subarrays FULL, readout FAST, groups
100, integration 1, exposures 36, resulting in a similar exposure time
of 10090 s.  The predictions for the JWST telescope are summarized in
Table~\ref{tab:JWST}. The general conclusion is that UCD progenitors
are detectable using JWST photometry with a $\approx 3$h exposure time
with promising values of $S/N$, as already suggested for GCs
progenitors by \cite{Renzini2017}.

\begin{table}[t] \centering \begin{tabular}{c | c | c | c | c  }
                instr. & filter & $z$ & $S/N$ CAN & $S/N$ MKDP \\ 
                \hline
                NIRCam & F115W &  3  & 47 & 194 \\\hline
                NIRCam & F200W &  3  & 47 & 215  \\\hline
                NIRCam & F480M &  3  & 7  & 52   \\\hline
                MIRI   & F1000W & 3  & 0.4 & 4   \\\hline
                \hline
                NIRCam & F115W &  6  & 14 & 76 \\\hline
                NIRCam & F200W &  6  & 15 & 83  \\\hline
                NIRCam & F480M &  6  & 2  & 16   \\\hline
                MIRI   & F1000W & 6  & 0.1 & 1   \\\hline
                \hline
                NIRCam & F115W &  9  & 6 & 36 \\\hline
                NIRCam & F200W &  9  & 7 & 42  \\\hline
                NIRCam & F480M &  9  & 1  & 7   \\\hline
                MIRI   & F1000W & 9  & 0.06 & 0.5   \\\hline
                
        \end{tabular} \centering \caption{
                Predictions for different filters
                of the JWST telescope. 
                All values are computed for an initial stellar mass of
                $10^8\,M_{\odot}$ and for reference the CAN IMF and the 
                MKDP IMF are considered. The $S/N$ values can be reached 
                within a total integration time of $\approx$ 3 hours.
                Section~\ref{sec:JWSTpreds} gives more details.
        }
        \label{tab:JWST}\end{table}

\section{Discussion}

\subsection{Where to look}
The star-formation-rate density typically peaks near the centre of a
galaxy.  This is evident in interacting galaxies
\citep{Joseph85,Norman87,Wright88,Dabringhausen2012}, while in
self-regulated galaxies the distribution of star-formation-rate
density may be more complex as a result of converging gas flows,
for example at the intersection points between a disc and bar. Generally
though, central regions are the most active in star formation activity
in star forming galaxies.  That the most massive clusters form near
the centres of galaxies where the SFR-density is highest is evident in
various star-bursting galaxies \citep{FM02,Dabringhausen2012}, in our
Galaxy \citep{Stolte14}, and also in young-cluster surveys of
individual galaxies \citep{Pflamm13}. Simulations of star forming galaxies lead to the same
result \citep{LiGnedin2017}.  Observationally it has been shown that the
most massive clusters form preferentially in galaxies with the highest 
star formation rate (SFR) \citep{Weidner2004, Randriamanakoto2013}. We can therefore
consider the following overall process: in the process of the
formation of massive galaxies, star formation would have been
spread throughout the merging proto-galactic gas clumps. The most
massive clusters, the proto-UCDs, would be forming in the deepest
potential wells of these, but would decouple from the hydrodynamics
once they became stellar systems. As the proto-galaxies merge to form
the massive central galaxies of galaxy clusters, many of such formed
UCDs would end up on orbits about the central galaxy, possibly with
the most massive UCDs within and near the centre of the galaxy.

We can therefore expect that the most massive UCDs, those that formed
monolithically (i.e. according to formation scenario~A), are to be
found in the innermost regions of forming central-dominant galaxies at
a high redshift.  Such galaxies have been deduced to form on a
timescale of and within less than a Gyr of the Big Bang with SFRs
larger than a few $10^3\,M_\odot/$yr \citep{Recchi2009}.  Under more
benign conditions, that is when the system-wide SFR is smaller as in
later interacting galaxies or the formation of less massive elliptical
galaxies within a few~Gyr of the Big Bang, UCDs may form also but are
more likely to be the mergers of star-cluster complexes.  Cases in
point are the Antennae galaxies, where such young complexes are evident
\citep{Kroupa1998,Fellhauer2002a}, and the Tadpole galaxy \cite{Kroupa2015}. 

Therefore, the best place to search for very massive UCDs is in the
inner regions of extreme star-bursts at very high redshift.  Elliptical galaxies and bulges in formation may also host young massive UCDs.  If
it were possible to observationally remove the gas and dust
obscuration, then such systems are likely to look like brilliantly lit
Christmas trees.

\subsection{Which IMF is to be expected if the formation scenario (A) or (B) is realized?} \label{sec:AorB}
If monolithic collapse (scenario~A) applies, then according to
\cite{Marks2012TH, Dabringhausen2012} top-heavy MKDP IMFs are expected, which
in turn lead to larger $M/L_V$ values at older ages (Figs. \ref{fig:MtoLV}
and \ref{fig:MtoLVLV}). At young ages, such objects can be as bright as 
quasars (Fig. \ref{fig:Lbol}).

In the case of the formation of
UCDs from merged cluster complexes (scenario~B), the IMF in each
sub-cluster would be closer to the canonical IMF. The IMF of a whole
object would thus be less top-heavy
\citep[e.g.][]{Pflamm2009,Weidner2013,Fontanot2017,Yan2017},  leading to
smaller $M/L_V$ ratios at older ages (Figs. \ref{fig:MtoLV}
        and \ref{fig:MtoLVLV}). They are significantly less luminous than monolithically-formed objects with MKDP IMFs (Fig. \ref{fig:Lbol}).
Therefore the realization of both scenarios
(A) and (B) in reality may lead to a spread of
$M/L_V$ values for present-day UCDs, which may be comparable to the observed 
spread. 

In the case of a bottom-heavy IMF, one might expect similar behaviour;
that is, if a UCD is created by a monolithic star-burst the IMF may be
more bottom-heavy than in the case of merged cluster complexes.
Thus, for monolithically-formed objects, the $M/L_V$ ratios are large at all ages (Fig. \ref{fig:MtoLV}), while they would be sub-luminous for their mass (Fig. \ref{fig:Lbol}).

All these different IMF cases can be distinguished best when the objects are younger 
than 100~Myr as their luminosities and colours will be most different (Fig. \ref{fig:McolD_MKDP}, Fig. \ref{fig:McolD}). Particularly useful observable diagnostics are provided by the colour-colour plots (Fig. \ref{fig:col_col}) and  the slope of the SED (Fig. \ref{fig:betaUV}).

\subsection{The implications of observed $M/L_V$--ratios} \label{Sec:MLv} 
The
elevated $M/L_V$ ratios observed for some UCDs can be caused by three, 
partially interconnected scenarios: (1) a variation of the IMF (top-heavy or
bottom-heavy), (2) the presence of a super-massive black hole (SMBH),
and (3) the presence of non-baryonic dark matter. Point (3) is
directly connected to formation scenario (D), which is, as already
mentioned, disfavoured for UCDs.  
As shown above, the variable IMF (scenario (1)) can explain the 
observed elevated $M/L_V$ values.  
The mass of a SMBH in scenario (2) required 
to explain the observed $M/L_V$ values, needs typically to be
10-15 per cent of the present-day UCD mass \citep{Mieske2013}. The presence
of SMBHs with such masses is indeed suggested or observationally confirmed at
least for a few UCDs \citep[e.g.][]{Seth2014, Janz2015,Ahn2017}. 

To address which scenario (variation of IMF, (1), or presence of SMBH,
(2)) is responsible for the elevated $M/L_V$ values in different
formation scenarios is not straightforward. In the case of formation
scenario (C), where UCDs are tidally stripped nuclei, there is an
observational connection with the presence of SMBHs. \cite{Graham2009}
found that the existence of a SMBH in a galactic nuclear cluster is
indeed frequently the case.  However, since the exact formation
mechanism of SMBHs is still discussed, it is not possible to constrain
the IMF of UCDs formed by scenario (C).  On the other hand, even
though in formation scenarios (A) and (B), where UCDs are cluster-like
objects and where the variable IMF is introduced to explain observed
$M/L_V$ values, the existence of a SMBH cannot be excluded.  The SMBH
can potentially be formed as a merger of dark remnants \citep[][Kroupa, et al,
in prep.]{Giersz2015}.  If a standard IMF is assumed, the mass of dark remnants
(BHs and neutron stars) represents approximately only 2 per cent of the
present-day mass of a system. 
For a top-heavy IMF this fraction can be significantly  higher.

\section{Conclusions} We investigated if observations  with
        upcoming observatories, with an emphasis on predictions for the
        JWST, may be able to discern the formation and evolution of UCDs
assuming that they are cluster-like objects which form by (A) single
monolithic collapse or (and) (B) by the merging of cluster
complexes. The primary area of interest is to find observable
        diagnostics that  may allow us to assess how the stellar IMF varies
        with physical conditions. The extreme star-bursts, which massive GCs
        and UCDs must have been at a high redshift, may be excellent test
        beds for this goal. For this purpose we compute the
time-dependent evolution of SEDs for different physical parameters and
mainly for different IMFs.  We test the top-heavy IMF, as parametrized
by \cite{Marks2012TH}, which predicts a top-heavy IMF for the case of
scenario~(A) and an IMF closer to the canonical IMF in the case of
scenario~(B). The bottom-heavy IMFs are implemented as a single power
law function with a slope $-2.3$ (Salpeter) and $-3$
\citep{vanDokkum2010}.

The main conclusions concerning the formation scenario for
UCDs and the stellar IMF variability can be summarized in five main
points: 
\begin{itemize} 
        \item The retention fraction of stellar remnants is near to 100\% for
        systems with birth masses larger than $10^7\,M_{\odot}$.
        
        \item We show that if UCD progenitors younger than
        $\approx 100\,\mathrm{Myr}$ are observed, their stellar IMF can be
        constrained and therefore also the formation scenario can be constrained by obtaining
        achievable measurements (e.g.\ absolute luminosity and supernova
        rate or an appropriate combination of colours and the value of
        $\beta_{UV}$).  UCD progenitors most likely located at redshifts 3-9
        have not been observed yet, however, according to our predictions we
        should be able to detect them even with current telescopes as they would
        appear like point sources with high, quasar-like
        luminosities. Computed exposure times for chosen JWST MIRI and
        NIRCam instruments are presented.  
        
        \item We also discuss degeneracies, which start appearing at ages
        $> 50\,\mathrm{Myr}$ as massive stars and evolve into dark remnants, and
        we reveal which information and constraints we can obtain from
        present-day UCDs. That is, the object's luminosity with a top-heavy
        IMF starts to be comparable with a UCD of the same (or even smaller)
        initial mass but with a canonical or bottom-heavy IMF. Therefore,
        within observational uncertainties, these cases might be
        indistinguishable on colour-magnitude or colour-colour diagrams.  Even
        $M/L_V$ becomes degenerate, however, for the majority of cases we
        should be able to separate a vDC IMF from the rest if the
        metallicity of the UCD is constrained reasonably well.
        
        \item If UCDs were formed with a top-heavy IMF ($\alpha_3<2.3$ with
        the most extreme case considered $\alpha_3=0.6$), their progenitors
        are extreme and very different from Galactic star formation
        regions. The UCD progenitors with initial stellar masses of
        $\approx 10^9\,\mathrm{M_{\odot}}$ would contain $\approx 10^7$ O
        stars in a region spanning not more than a few~pc. This drives a
        tremendous luminosity, very high SN II rates, and also poses the further question as to how the strong radiation field influences the state and evolution of other stars and thus the IMF especially at the low-mass end \citep[e.g.][]{Kroupa2003}.
        
        \item Interestingly, we have found evidence that some of the observed
        quasars have photometric properties of very young UCD models with
        top-heavy IMFs. This may suggest that some quasars at high redshift may actually be
        very massive UCDs with ages $<10\,$Myr.  This needs further study
        though, for example\ by quantifications of SEDs. One method to help
        identify true UCDs with top-heavy IMFs would be to monitor their
        luminosities. Since core-collapse supernovae will be common in such
        systems, exploding at a rate of more than one per year, the
        luminosity of such a UCD ought to show increases by $\approx 0.1-10$
        per cent (depending on star formation history, IMF, and initial
        stellar mass) over a timescale of a few months up to a few dozen
        times a year. Young UCDs should thus be time-variable.

        \item Groups of very young UCDs, if found at high redshift, may be
        indicating the assembly of the inner regions of galaxy clusters
        \citep[see also][]{Schulz2016}: the assembly timescale is
        $10^2\,$Myr, being of the order of the dynamical timescale.  The
        seeds of the most massive galaxies in the centre of galaxy clusters
        probably had a very clustered formation of UCD-mass objects that
        created today's giant ellipticals and brightest cluster galaxies.
        Thus, during the assembly of the inner region of galaxy clusters, we
        would expect generations of quasar-like UCDs, each with a high
        luminosity and lifetime of about $10\,$Myr, forming such that the
        overall lifetime of the UCD-active epoch would be about
        $10^2\,$Myr. This is comparable to the lifetime of quasars, adding
        to the similarity in photometric properties noted above.
        
        \item The majority of ultra-massive very young UCDs, which look
        comparable to quasars, are therefore likely to form in the central
        region of the star-bursts from which the present-day central
        dominant elliptical galaxies emerge. But such UCDs will not
        be observable today as they are likely to sink to the centres of the
        elliptical galaxies through dynamical friction \cite{Bekki2010}.
        
\end{itemize} 

To gain more firm conclusions, individual cases of observed UCDs need
to be considered taking all observational constraints into account.  To
disentangle degeneracies that arise mainly with age, new data
reporting UCDs younger than $\approx 100\,\mathrm{Myr}$ are needed. We
would like to emphasize here that no such objects have been
observationally confirmed yet.

\begin{acknowledgements}
        We thank the referee and Holger Baumgardt for useful comments 
                that helped to improve this manuscript.
        TJ was supported by Charles University in Prague through grant
        SVV-260441 and through a stipend from the SPODYR group at the
        University of Bonn.  TJ, PK, and KB thank the DAAD (grant 57212729
        "Galaxy formation with a variable stellar initial mass function”)
        for funding exchange visits. We would like to acknowledge the use
        of Python (G. van Rossum, Python tutorial, Technical Report
        CS-R9526, Centrum voor Wiskunde en Informatica (CWI), Amsterdam, May
        1995).  Apart from standard Python libraries, we used pyPegase from
        Colin Jacobs.  We also acknowledge discussions with Christopher Tout
        and many useful contributions seen during the ImBaSe 2017 conference
        in Garching.
\end{acknowledgements}

\bibliographystyle{aa}   
\bibliography{library}

\begin{thebibliography}{114}
\expandafter\ifx\csname natexlab\endcsname\relax\def\natexlab#1{#1}\fi

\bibitem[{{Adams} \& {Laughlin}(1996)}]{Adams1996}
{Adams}, F.~C. \& {Laughlin}, G. 1996, \apj, 468, 586

\bibitem[{Ahn {et~al.}(2017)Ahn, Seth, den Brok, Strader, Baumgardt, van~den
  Bosch, Chilingarian, Frank, Hilker, McDermid, Mieske, Romanowsky, Spitler,
  Brodie, Neumayer, \& Walsh}]{Ahn2017}
Ahn, C.~P., Seth, A.~C., den Brok, M., {et~al.} 2017, The Astrophysical
  Journal, 839, 72

\bibitem[{{Anderson} {et~al.}(2011){Anderson}, {Habergham}, \&
  {James}}]{Anderson2011}
{Anderson}, J.~P., {Habergham}, S.~M., \& {James}, P.~A. 2011, \mnras, 416, 567

\bibitem[{Balbinot \& Gieles(2017)}]{Balbinot2017}
Balbinot, E. \& Gieles, M. 2017, arXiv.org, arXiv:1702.02543

\bibitem[{{Banerjee}(2017)}]{Banerjee2017}
{Banerjee}, S. 2017, \mnras, 467, 524

\bibitem[{{Bastian} {et~al.}(2010){Bastian}, {Covey}, \& {Meyer}}]{Bastian2010}
{Bastian}, N., {Covey}, K.~R., \& {Meyer}, M.~R. 2010, \araa, 48, 339

\bibitem[{{Bastian} \& {Strader}(2014)}]{Bastian2014}
{Bastian}, N. \& {Strader}, J. 2014, \mnras, 443, 3594

\bibitem[{Baumgardt \& Sollima(2017)}]{Baumgardt2017}
Baumgardt, H. \& Sollima, A. 2017, arXiv.org, arXiv:1708.09530

\bibitem[{{Bekki}(2010)}]{Bekki2010}
{Bekki}, K. 2010, \mnras, 401, 2753

\bibitem[{{Bekki} {et~al.}(2001){Bekki}, {Couch}, \& {Drinkwater}}]{Bekki2001}
{Bekki}, K., {Couch}, W.~J., \& {Drinkwater}, M.~J. 2001, \apjl, 552, L105

\bibitem[{{Bekki} {et~al.}(2003){Bekki}, {Couch}, {Drinkwater}, \&
  {Shioya}}]{Bekki2003}
{Bekki}, K., {Couch}, W.~J., {Drinkwater}, M.~J., \& {Shioya}, Y. 2003, \mnras,
  344, 399

\bibitem[{{Bertelli Motta} {et~al.}(2016){Bertelli Motta}, {Clark}, {Glover},
  {Klessen}, \& {Pasquali}}]{Bertelli2016}
{Bertelli Motta}, C., {Clark}, P.~C., {Glover}, S.~C.~O., {Klessen}, R.~S., \&
  {Pasquali}, A. 2016, \mnras, 462, 4171

\bibitem[{Brinkmann {et~al.}(2017)Brinkmann, Banerjee, Motwani, \&
  Kroupa}]{Brinkmann2017}
Brinkmann, N., Banerjee, S., Motwani, B., \& Kroupa, P. 2017, Astronomy {\&}
  Astrophysics, 600, A49

\bibitem[{{Brodie} {et~al.}(2011){Brodie}, {Romanowsky}, {Strader}, \&
  {Forbes}}]{Brodie2011}
{Brodie}, J.~P., {Romanowsky}, A.~J., {Strader}, J., \& {Forbes}, D.~A. 2011,
  \aj, 142, 199

\bibitem[{{Br{\"u}ns} {et~al.}(2011){Br{\"u}ns}, {Kroupa}, {Fellhauer}, {Metz},
  \& {Assmann}}]{Bruens2011}
{Br{\"u}ns}, R.~C., {Kroupa}, P., {Fellhauer}, M., {Metz}, M., \& {Assmann}, P.
  2011, \aap, 529, A138

\bibitem[{{Chabrier} {et~al.}(2014){Chabrier}, {Hennebelle}, \&
  {Charlot}}]{Chabrier2014}
{Chabrier}, G., {Hennebelle}, P., \& {Charlot}, S. 2014, \apj, 796, 75

\bibitem[{{Chiboucas} {et~al.}(2011){Chiboucas}, {Tully}, {Marzke},
  {Phillipps}, {Price}, {Peng}, {Trentham}, {Carter}, \&
  {Hammer}}]{Chiboucas2011}
{Chiboucas}, K., {Tully}, R.~B., {Marzke}, R.~O., {et~al.} 2011, \apj, 737, 86

\bibitem[{{Chilingarian} {et~al.}(2008){Chilingarian}, {Cayatte}, \&
  {Bergond}}]{Chilingarian2008}
{Chilingarian}, I.~V., {Cayatte}, V., \& {Bergond}, G. 2008, \mnras, 390, 906

\bibitem[{{C{\^o}t{\'e}} {et~al.}(2006){C{\^o}t{\'e}}, {Piatek}, {Ferrarese},
  {Jord{\'a}n}, {Merritt}, {Peng}, {Ha{\c s}egan}, {Blakeslee}, {Mei}, {West},
  {Milosavljevi{\'c}}, \& {Tonry}}]{Cote2006}
{C{\^o}t{\'e}}, P., {Piatek}, S., {Ferrarese}, L., {et~al.} 2006, \apjs, 165,
  57

\bibitem[{{Dabringhausen} {et~al.}(2010){Dabringhausen}, {Fellhauer}, \&
  {Kroupa}}]{Dabringhausen2010}
{Dabringhausen}, J., {Fellhauer}, M., \& {Kroupa}, P. 2010, \mnras, 403, 1054

\bibitem[{{Dabringhausen} {et~al.}(2008){Dabringhausen}, {Hilker}, \&
  {Kroupa}}]{Dabringhausen2008}
{Dabringhausen}, J., {Hilker}, M., \& {Kroupa}, P. 2008, \mnras, 386, 864

\bibitem[{Dabringhausen {et~al.}(2009)Dabringhausen, Kroupa, \&
  Baumgardt}]{Dabringhausen2009}
Dabringhausen, J., Kroupa, P., \& Baumgardt, H. 2009, Monthly Notices of the
  Royal Astronomical Society, 394, 1529

\bibitem[{{Dabringhausen} {et~al.}(2012){Dabringhausen}, {Kroupa},
  {Pflamm-Altenburg}, \& {Mieske}}]{Dabringhausen2012}
{Dabringhausen}, J., {Kroupa}, P., {Pflamm-Altenburg}, J., \& {Mieske}, S.
  2012, \apj, 747, 72

\bibitem[{Dib {et~al.}(2007)Dib, Kim, \& Shadmehri}]{Dib2007}
Dib, S., Kim, J., \& Shadmehri, M. 2007, Monthly Notices of the Royal
  Astronomical Society: Letters, 381, L40

\bibitem[{{Drinkwater} {et~al.}(2004){Drinkwater}, {Gregg}, {Couch},
  {Ferguson}, {Hilker}, {Jones}, {Karick}, \& {Phillipps}}]{Drinkwater2004}
{Drinkwater}, M.~J., {Gregg}, M.~D., {Couch}, W.~J., {et~al.} 2004, \pasa, 21,
  375

\bibitem[{{Drinkwater} {et~al.}(2003){Drinkwater}, {Gregg}, {Hilker}, {Bekki},
  {Couch}, {Ferguson}, {Jones}, \& {Phillipps}}]{Drinkwater2003}
{Drinkwater}, M.~J., {Gregg}, M.~D., {Hilker}, M., {et~al.} 2003, \nat, 423,
  519

\bibitem[{{Drinkwater} {et~al.}(2000){Drinkwater}, {Jones}, {Gregg}, \&
  {Phillipps}}]{Drinkwater2000}
{Drinkwater}, M.~J., {Jones}, J.~B., {Gregg}, M.~D., \& {Phillipps}, S. 2000,
  \pasa, 17, 227

\bibitem[{{Dunlop} {et~al.}(2003){Dunlop}, {McLure}, {Kukula}, {Baum}, {O'Dea},
  \& {Hughes}}]{Dunlop2003}
{Dunlop}, J.~S., {McLure}, R.~J., {Kukula}, M.~J., {et~al.} 2003, \mnras, 340,
  1095

\bibitem[{{Dunlop} {et~al.}(1993){Dunlop}, {Taylor}, {Hughes}, \&
  {Robson}}]{Dunlop1993}
{Dunlop}, J.~S., {Taylor}, G.~L., {Hughes}, D.~H., \& {Robson}, E.~I. 1993,
  \mnras, 264, 455

\bibitem[{{Elmegreen}(2004)}]{Elmegreen2004}
{Elmegreen}, B.~G. 2004, \mnras, 354, 367

\bibitem[{{Evstigneeva} {et~al.}(2008){Evstigneeva}, {Drinkwater}, {Peng},
  {Hilker}, {De Propris}, {Jones}, {Phillipps}, {Gregg}, \&
  {Karick}}]{Evstigneeva2008}
{Evstigneeva}, E.~A., {Drinkwater}, M.~J., {Peng}, C.~Y., {et~al.} 2008, \aj,
  136, 461

\bibitem[{{Fellhauer} \& {Kroupa}(2002{\natexlab{a}})}]{Fellhauer2002b}
{Fellhauer}, M. \& {Kroupa}, P. 2002{\natexlab{a}}, \apss, 281, 355

\bibitem[{{Fellhauer} \& {Kroupa}(2002{\natexlab{b}})}]{Fellhauer2002a}
{Fellhauer}, M. \& {Kroupa}, P. 2002{\natexlab{b}}, \mnras, 330, 642

\bibitem[{{Ferrarese} \& {Merritt}(2002)}]{FM02}
{Ferrarese}, L. \& {Merritt}, D. 2002, Phys.~World, 15N6, 41

\bibitem[{{Fioc} \& {Rocca-Volmerange}(1997)}]{Fioc1997}
{Fioc}, M. \& {Rocca-Volmerange}, B. 1997, \aap, 326, 950

\bibitem[{{Fontanot} {et~al.}(2017){Fontanot}, {De Lucia}, {Hirschmann},
  {Bruzual}, {Charlot}, \& {Zibetti}}]{Fontanot2017}
{Fontanot}, F., {De Lucia}, G., {Hirschmann}, M., {et~al.} 2017, \mnras, 464,
  3812

\bibitem[{{Forbes} {et~al.}(2008){Forbes}, {Lasky}, {Graham}, \&
  {Spitler}}]{Forbes2008}
{Forbes}, D.~A., {Lasky}, P., {Graham}, A.~W., \& {Spitler}, L. 2008, \mnras,
  389, 1924

\bibitem[{{Frank} {et~al.}(2011){Frank}, {Hilker}, {Mieske}, {Baumgardt},
  {Grebel}, \& {Infante}}]{Frank2011}
{Frank}, M.~J., {Hilker}, M., {Mieske}, S., {et~al.} 2011, \mnras, 414, L70

\bibitem[{{Gal-Yam}(2012)}]{Gal-Yam2012}
{Gal-Yam}, A. 2012, Science, 337, 927

\bibitem[{Giersz {et~al.}(2015)Giersz, Leigh, Hypki, L{\"u}tzgendorf, \&
  Askar}]{Giersz2015}
Giersz, M., Leigh, N., Hypki, A., L{\"u}tzgendorf, N., \& Askar, A. 2015,
  \mnras, 454, 3150

\bibitem[{Glazebrook {et~al.}(2017)Glazebrook, Schreiber, Labb{\'e},
  Nanayakkara, Kacprzak, Oesch, Papovich, Spitler, Straatman, Tran, \&
  Yuan}]{Glazebrook2017}
Glazebrook, K., Schreiber, C., Labb{\'e}, I., {et~al.} 2017, Nature, 544, 71

\bibitem[{{Goerdt} {et~al.}(2008){Goerdt}, {Moore}, {Kazantzidis}, {Kaufmann},
  {Macci{\`o}}, \& {Stadel}}]{Goerdt2008}
{Goerdt}, T., {Moore}, B., {Kazantzidis}, S., {et~al.} 2008, \mnras, 385, 2136

\bibitem[{{Graham} \& {Spitler}(2009)}]{Graham2009}
{Graham}, A.~W. \& {Spitler}, L.~R. 2009, \mnras, 397, 2148

\bibitem[{{Greggio} \& {Renzini}(1983)}]{Greggio1983}
{Greggio}, L. \& {Renzini}, A. 1983, \aap, 118, 217

\bibitem[{{Hilker}(2009)}]{Hilker2009}
{Hilker}, M. 2009, {UCDs - A Mixed Bag of Objects}, ed. T.~{Richtler} \&
  S.~{Larsen}, 51

\bibitem[{{Hilker} {et~al.}(1999){Hilker}, {Infante}, {Vieira},
  {Kissler-Patig}, \& {Richtler}}]{Hilker1999}
{Hilker}, M., {Infante}, L., {Vieira}, G., {Kissler-Patig}, M., \& {Richtler},
  T. 1999, \aaps, 134, 75

\bibitem[{{Janz} {et~al.}(2015){Janz}, {Forbes}, {Norris}, {Strader}, {Penny},
  {Fagioli}, \& {Romanowsky}}]{Janz2015}
{Janz}, J., {Forbes}, D.~A., {Norris}, M.~A., {et~al.} 2015, \mnras, 449, 1716

\bibitem[{{Joseph} \& {Wright}(1985)}]{Joseph85}
{Joseph}, R.~D. \& {Wright}, G.~S. 1985, \mnras, 214, 87

\bibitem[{{Kroupa}(1998)}]{Kroupa1998}
{Kroupa}, P. 1998, \mnras, 300, 200

\bibitem[{{Kroupa}(2001)}]{Kroupa2001}
{Kroupa}, P. 2001, \mnras, 322, 231

\bibitem[{Kroupa(2015)}]{Kroupa2015}
Kroupa, P. 2015, Canadian Journal of Physics, 93, 169

\bibitem[{{Kroupa} \& {Bouvier}(2003)}]{Kroupa2003}
{Kroupa}, P. \& {Bouvier}, J. 2003, \mnras, 346, 369

\bibitem[{{Kroupa} {et~al.}(2013){Kroupa}, {Weidner}, {Pflamm-Altenburg},
  {Thies}, {Dabringhausen}, {Marks}, \& {Maschberger}}]{Kroupa2013}
{Kroupa}, P., {Weidner}, C., {Pflamm-Altenburg}, J., {et~al.} 2013, {The
  Stellar and Sub-Stellar Initial Mass Function of Simple and Composite
  Populations}, 115

\bibitem[{{Lamers} {et~al.}(2005){Lamers}, {Gieles}, {Bastian}, {Baumgardt},
  {Kharchenko}, \& {Portegies Zwart}}]{Lamers2005}
{Lamers}, H.~J.~G.~L.~M., {Gieles}, M., {Bastian}, N., {et~al.} 2005, \aap,
  441, 117

\bibitem[{{Larson}(1998)}]{Larsen1998}
{Larson}, R.~B. 1998, \mnras, 301, 569

\bibitem[{{Leitherer} {et~al.}(1999){Leitherer}, {Schaerer}, {Goldader},
  {Delgado}, {Robert}, {Kune}, {de Mello}, {Devost}, \&
  {Heckman}}]{Leitherer1999}
{Leitherer}, C., {Schaerer}, D., {Goldader}, J.~D., {et~al.} 1999, \apjs, 123,
  3

\bibitem[{Li {et~al.}(2017)Li, Gnedin, Gnedin, Meng, Semenov, \&
  Kravtsov}]{LiGnedin2017}
Li, H., Gnedin, O.~Y., Gnedin, N.~Y., {et~al.} 2017, The Astrophysical Journal,
  834, 69

\bibitem[{{Liptai} {et~al.}(2017){Liptai}, {Price}, {Wurster}, \&
  {Bate}}]{Liptai2017}
{Liptai}, D., {Price}, D.~J., {Wurster}, J., \& {Bate}, M.~R. 2017, \mnras,
  465, 105

\bibitem[{{Longmore}(2015)}]{Longmore2015}
{Longmore}, S.~N. 2015, \mnras, 448, L62

\bibitem[{{Lonsdale} {et~al.}(2006){Lonsdale}, {Diamond}, {Thrall}, {Smith}, \&
  {Lonsdale}}]{Lonsdale2006}
{Lonsdale}, C.~J., {Diamond}, P.~J., {Thrall}, H., {Smith}, H.~E., \&
  {Lonsdale}, C.~J. 2006, \apj, 647, 185

\bibitem[{Lyman {et~al.}(2016)Lyman, Bersier, James, Mazzali, Eldridge, Fraser,
  \& Pian}]{Lyman2016}
Lyman, J.~D., Bersier, D., James, P.~A., {et~al.} 2016, Monthly Notices of the
  Royal Astronomical Society, 457, 328

\bibitem[{{Maoz}(2008)}]{Maoz2008}
{Maoz}, D. 2008, \mnras, 384, 267

\bibitem[{{Marks} \& {Kroupa}(2012)}]{Marks2012}
{Marks}, M. \& {Kroupa}, P. 2012, \aap, 543, A8

\bibitem[{{Marks} {et~al.}(2012){Marks}, {Kroupa}, {Dabringhausen}, \&
  {Pawlowski}}]{Marks2012TH}
{Marks}, M., {Kroupa}, P., {Dabringhausen}, J., \& {Pawlowski}, M.~S. 2012,
  \mnras, 422, 2246

\bibitem[{{Marks} {et~al.}(2011){Marks}, {Kroupa}, \& {Oh}}]{Marks2011}
{Marks}, M., {Kroupa}, P., \& {Oh}, S. 2011, \mnras, 417, 1684

\bibitem[{{Matteucci} \& {Greggio}(1986)}]{Matteucci1986}
{Matteucci}, F. \& {Greggio}, L. 1986, \aap, 154, 279

\bibitem[{{Megeath} {et~al.}(2016){Megeath}, {Gutermuth}, {Muzerolle},
  {Kryukova}, {Hora}, {Allen}, {Flaherty}, {Hartmann}, {Myers}, {Pipher},
  {Stauffer}, {Young}, \& {Fazio}}]{Megeath2016}
{Megeath}, S.~T., {Gutermuth}, R., {Muzerolle}, J., {et~al.} 2016, \aj, 151, 5

\bibitem[{{Mieske} {et~al.}(2008){Mieske}, {Dabringhausen}, {Kroupa}, {Hilker},
  \& {Baumgardt}}]{Mieske2008}
{Mieske}, S., {Dabringhausen}, J., {Kroupa}, P., {Hilker}, M., \& {Baumgardt},
  H. 2008, Astronomische Nachrichten, 329, 964

\bibitem[{{Mieske} {et~al.}(2013){Mieske}, {Frank}, {Baumgardt},
  {L{\"u}tzgendorf}, {Neumayer}, \& {Hilker}}]{Mieske2013}
{Mieske}, S., {Frank}, M.~J., {Baumgardt}, H., {et~al.} 2013, \aap, 558, A14

\bibitem[{{Mieske} {et~al.}(2002){Mieske}, {Hilker}, \& {Infante}}]{Mieske2002}
{Mieske}, S., {Hilker}, M., \& {Infante}, L. 2002, \aap, 383, 823

\bibitem[{{Mieske} {et~al.}(2007){Mieske}, {Hilker}, {Jord{\'a}n}, {Infante},
  \& {Kissler-Patig}}]{Mieske2007}
{Mieske}, S., {Hilker}, M., {Jord{\'a}n}, A., {Infante}, L., \&
  {Kissler-Patig}, M. 2007, \aap, 472, 111

\bibitem[{{Mieske} {et~al.}(2012){Mieske}, {Hilker}, \& {Misgeld}}]{Mieske2012}
{Mieske}, S., {Hilker}, M., \& {Misgeld}, I. 2012, \aap, 537, A3

\bibitem[{{Mieske} \& {Kroupa}(2008)}]{Mieske2008b}
{Mieske}, S. \& {Kroupa}, P. 2008, \apj, 677, 276

\bibitem[{{Misgeld} {et~al.}(2011){Misgeld}, {Mieske}, {Hilker}, {Richtler},
  {Georgiev}, \& {Schuberth}}]{Misgeld2011}
{Misgeld}, I., {Mieske}, S., {Hilker}, M., {et~al.} 2011, \aap, 531, A4

\bibitem[{Mortlock {et~al.}(2011)Mortlock, Warren, Venemans, Patel, Hewett,
  McMahon, Simpson, Theuns, Gonz{\'a}les-Solares, Adamson, Dye, Hambly, Hirst,
  Irwin, Kuiper, Lawrence, \& R{\"o}ttgering}]{Mortlock2011}
Mortlock, D.~J., Warren, S.~J., Venemans, B.~P., {et~al.} 2011, Nature, 474,
  616

\bibitem[{{Murray}(2009)}]{Murray2009}
{Murray}, N. 2009, \apj, 691, 946

\bibitem[{{Norman}(1987)}]{Norman87}
{Norman}, C.~A. 1987, in NASA Conference Publication, Vol. 2466, NASA
  Conference Publication, ed. C.~J. {Lonsdale Persson}

\bibitem[{{Oh} {et~al.}(1995){Oh}, {Lin}, \& {Aarseth}}]{Oh1995}
{Oh}, K.~S., {Lin}, D.~N.~C., \& {Aarseth}, S.~J. 1995, \apj, 442, 142

\bibitem[{{Papadopoulos}(2010)}]{Papadopoulos2010}
{Papadopoulos}, P.~P. 2010, \apj, 720, 226

\bibitem[{{Pejcha} \& {Thompson}(2015)}]{Pejcha2015}
{Pejcha}, O. \& {Thompson}, T.~A. 2015, \apj, 801, 90

\bibitem[{{Penny} {et~al.}(2012){Penny}, {Forbes}, \& {Conselice}}]{Penny2012}
{Penny}, S.~J., {Forbes}, D.~A., \& {Conselice}, C.~J. 2012, \mnras, 422, 885

\bibitem[{{Peuten} {et~al.}(2016){Peuten}, {Zocchi}, {Gieles}, {Gualandris}, \&
  {H{\'e}nault-Brunet}}]{Peuten2016}
{Peuten}, M., {Zocchi}, A., {Gieles}, M., {Gualandris}, A., \&
  {H{\'e}nault-Brunet}, V. 2016, \mnras, 462, 2333

\bibitem[{{Pfeffer} \& {Baumgardt}(2013)}]{Pfeffer2013}
{Pfeffer}, J. \& {Baumgardt}, H. 2013, \mnras, 433, 1997

\bibitem[{{Pfeffer} {et~al.}(2014){Pfeffer}, {Griffen}, {Baumgardt}, \&
  {Hilker}}]{Pfeffer2014}
{Pfeffer}, J., {Griffen}, B.~F., {Baumgardt}, H., \& {Hilker}, M. 2014, \mnras,
  444, 3670

\bibitem[{{Pfeffer} {et~al.}(2016){Pfeffer}, {Hilker}, {Baumgardt}, \&
  {Griffen}}]{Pfeffer2016}
{Pfeffer}, J., {Hilker}, M., {Baumgardt}, H., \& {Griffen}, B.~F. 2016, \mnras,
  458, 2492

\bibitem[{{Pflamm-Altenburg} {et~al.}(2013){Pflamm-Altenburg},
  {Gonz{\'a}lez-L{\'o}pezlira}, \& {Kroupa}}]{Pflamm13}
{Pflamm-Altenburg}, J., {Gonz{\'a}lez-L{\'o}pezlira}, R.~A., \& {Kroupa}, P.
  2013, \mnras, 435, 2604

\bibitem[{{Pflamm-Altenburg} {et~al.}(2009){Pflamm-Altenburg}, {Weidner}, \&
  {Kroupa}}]{Pflamm2009}
{Pflamm-Altenburg}, J., {Weidner}, C., \& {Kroupa}, P. 2009, \mnras, 395, 394

\bibitem[{{Planck Collaboration} {et~al.}(2016{\natexlab{a}}){Planck
  Collaboration}, {Adam}, {Aghanim}, {Ashdown}, {Aumont}, {Baccigalupi},
  {Ballardini}, {Banday}, {Barreiro}, {Bartolo}, {Basak}, {Battye}, {Benabed},
  {Bernard}, {Bersanelli}, {Bielewicz}, {Bock}, {Bonaldi}, {Bonavera}, {Bond},
  {Borrill}, {Bouchet}, {Boulanger}, {Bucher}, {Burigana}, {Calabrese},
  {Cardoso}, {Carron}, {Chiang}, {Colombo}, {Combet}, {Comis}, {Couchot},
  {Coulais}, {Crill}, {Curto}, {Cuttaia}, {Davis}, {de Bernardis}, {de Rosa},
  {de Zotti}, {Delabrouille}, {Di Valentino}, {Dickinson}, {Diego}, {Dor{\'e}},
  {Douspis}, {Ducout}, {Dupac}, {Elsner}, {En{\ss}lin}, {Eriksen}, {Falgarone},
  {Fantaye}, {Finelli}, {Forastieri}, {Frailis}, {Fraisse}, {Franceschi},
  {Frolov}, {Galeotta}, {Galli}, {Ganga}, {G{\'e}nova-Santos}, {Gerbino},
  {Ghosh}, {Gonz{\'a}lez-Nuevo}, {G{\'o}rski}, {Gruppuso}, {Gudmundsson},
  {Hansen}, {Helou}, {Henrot-Versill{\'e}}, {Herranz}, {Hivon}, {Huang},
  {Ili{\'c}}, {Jaffe}, {Jones}, {Keih{\"a}nen}, {Keskitalo}, {Kisner}, {Knox},
  {Krachmalnicoff}, {Kunz}, {Kurki-Suonio}, {Lagache}, {L{\"a}hteenm{\"a}ki},
  {Lamarre}, {Langer}, {Lasenby}, {Lattanzi}, {Lawrence}, {Le Jeune},
  {Levrier}, {Lewis}, {Liguori}, {Lilje}, {L{\'o}pez-Caniego}, {Ma},
  {Mac{\'{\i}}as-P{\'e}rez}, {Maggio}, {Mangilli}, {Maris}, {Martin},
  {Mart{\'{\i}}nez-Gonz{\'a}lez}, {Matarrese}, {Mauri}, {McEwen}, {Meinhold},
  {Melchiorri}, {Mennella}, {Migliaccio}, {Miville-Desch{\^e}nes}, {Molinari},
  {Moneti}, {Montier}, {Morgante}, {Moss}, {Naselsky}, {Natoli}, {Oxborrow},
  {Pagano}, {Paoletti}, {Partridge}, {Patanchon}, {Patrizii}, {Perdereau},
  {Perotto}, {Pettorino}, {Piacentini}, {Plaszczynski}, {Polastri}, {Polenta},
  {Puget}, {Rachen}, {Racine}, {Reinecke}, {Remazeilles}, {Renzi}, {Rocha},
  {Rossetti}, {Roudier}, {Rubi{\~n}o-Mart{\'{\i}}n}, {Ruiz-Granados},
  {Salvati}, {Sandri}, {Savelainen}, {Scott}, {Sirri}, {Sunyaev}, {Suur-Uski},
  {Tauber}, {Tenti}, {Toffolatti}, {Tomasi}, {Tristram}, {Trombetti},
  {Valiviita}, {Van Tent}, {Vielva}, {Villa}, {Vittorio}, {Wandelt}, {Wehus},
  {White}, {Zacchei}, \& {Zonca}}]{Planck2016b}
{Planck Collaboration}, {Adam}, R., {Aghanim}, N., {et~al.} 2016{\natexlab{a}},
  \aap, 596, A108

\bibitem[{{Planck Collaboration} {et~al.}(2016{\natexlab{b}}){Planck
  Collaboration}, {Ade}, {Aghanim}, {Arnaud}, {Ashdown}, {Aumont},
  {Baccigalupi}, {Banday}, {Barreiro}, {Bartlett}, \& et~al.}]{Planck2016}
{Planck Collaboration}, {Ade}, P.~A.~R., {Aghanim}, N., {et~al.}
  2016{\natexlab{b}}, \aap, 594, A13

\bibitem[{{Price} {et~al.}(2009){Price}, {Phillipps}, {Huxor}, {Trentham},
  {Ferguson}, {Marzke}, {Hornschemeier}, {Goudfrooij}, {Hammer}, {Tully},
  {Chiboucas}, {Smith}, {Carter}, {Merritt}, {Balcells}, {Erwin}, \&
  {Puzia}}]{Price2009}
{Price}, J., {Phillipps}, S., {Huxor}, A., {et~al.} 2009, \mnras, 397, 1816

\bibitem[{{Randriamanakoto} {et~al.}(2013){Randriamanakoto}, {Escala},
  {V{\"a}is{\"a}nen}, {Kankare}, {Kotilainen}, {Mattila}, \&
  {Ryder}}]{Randriamanakoto2013}
{Randriamanakoto}, Z., {Escala}, A., {V{\"a}is{\"a}nen}, P., {et~al.} 2013,
  \apjl, 775, L38

\bibitem[{{Recchi} {et~al.}(2009){Recchi}, {Calura}, \& {Kroupa}}]{Recchi2009}
{Recchi}, S., {Calura}, F., \& {Kroupa}, P. 2009, \aap, 499, 711

\bibitem[{{Renaud} {et~al.}(2015){Renaud}, {Bournaud}, \& {Duc}}]{Renaud2015}
{Renaud}, F., {Bournaud}, F., \& {Duc}, P.-A. 2015, \mnras, 446, 2038

\bibitem[{Renzini(2017)}]{Renzini2017}
Renzini, A. 2017, Monthly Notices of the Royal Astronomical Society: Letters,
  469, L63

\bibitem[{Renzini {et~al.}(2015)Renzini, D'Antona, Cassisi, King, Milone,
  Ventura, Anderson, Bedin, Bellini, Brown, Piotto, van~der Marel, Barbuy,
  Dalessandro, Hidalgo, Marino, Ortolani, Salaris, \& Sarajedini}]{Renzini2015}
Renzini, A., D'Antona, F., Cassisi, S., {et~al.} 2015, Monthly Notices of the
  Royal Astronomical Society, 454, 4197

\bibitem[{Romano {et~al.}(2017)Romano, Matteucci, Zhang, Papadopoulos, \&
  Ivison}]{Romano2017}
Romano, D., Matteucci, F., Zhang, Z.~Y., Papadopoulos, P.~P., \& Ivison, R.~J.
  2017, Monthly Notices of the Royal Astronomical Society, 470, 401

\bibitem[{{Schulz} {et~al.}(2016){Schulz}, {Hilker}, {Kroupa}, \&
  {Pflamm-Altenburg}}]{Schulz2016}
{Schulz}, C., {Hilker}, M., {Kroupa}, P., \& {Pflamm-Altenburg}, J. 2016, \aap,
  594, A119

\bibitem[{{Schulz} {et~al.}(2015){Schulz}, {Pflamm-Altenburg}, \&
  {Kroupa}}]{Schulz2015}
{Schulz}, C., {Pflamm-Altenburg}, J., \& {Kroupa}, P. 2015, \aap, 582, A93

\bibitem[{{Seth} {et~al.}(2014){Seth}, {van den Bosch}, {Mieske}, {Baumgardt},
  {Brok}, {Strader}, {Neumayer}, {Chilingarian}, {Hilker}, {McDermid},
  {Spitler}, {Brodie}, {Frank}, \& {Walsh}}]{Seth2014}
{Seth}, A.~C., {van den Bosch}, R., {Mieske}, S., {et~al.} 2014, \nat, 513, 398

\bibitem[{{Smith} \& {Lucey}(2013)}]{Smith2013}
{Smith}, R.~J. \& {Lucey}, J.~R. 2013, \mnras, 434, 1964

\bibitem[{{Souchay} {et~al.}(2015){Souchay}, {Andrei}, {Barache}, {Kalewicz},
  {Gattano}, {Coelho}, {Taris}, {Bouquillon}, \& {Becker}}]{Souchay2015}
{Souchay}, J., {Andrei}, A.~H., {Barache}, C., {et~al.} 2015, \aap, 583, A75

\bibitem[{{Stanway} {et~al.}(2016){Stanway}, {Eldridge}, \&
  {Becker}}]{Stanway2016}
{Stanway}, E.~R., {Eldridge}, J.~J., \& {Becker}, G.~D. 2016, \mnras, 456, 485

\bibitem[{{Stolte} {et~al.}(2014){Stolte}, {Hu{\ss}mann}, {Morris}, {Ghez},
  {Lu}, {Clarkson}, \& {Habibi}}]{Stolte14}
{Stolte}, A., {Hu{\ss}mann}, B., {Morris}, M.~R., {et~al.} 2014, \apj, 789, 115

\bibitem[{{Terlevich} \& {Boyle}(1993)}]{Terlevich1993}
{Terlevich}, R.~J. \& {Boyle}, B.~J. 1993, \mnras, 262, 491

\bibitem[{{Thielemann} {et~al.}(1986){Thielemann}, {Nomoto}, \&
  {Yokoi}}]{Thielemann1986}
{Thielemann}, F.-K., {Nomoto}, K., \& {Yokoi}, K. 1986, \aap, 158, 17

\bibitem[{Thies {et~al.}(2015)Thies, Pflamm-Altenburg, Kroupa, \&
  Marks}]{Thies2015}
Thies, I., Pflamm-Altenburg, J., Kroupa, P., \& Marks, M. 2015, The
  Astrophysical Journal, 800, 72

\bibitem[{{Thomas} {et~al.}(2008){Thomas}, {Drinkwater}, \&
  {Evstigneeva}}]{Thomas2008}
{Thomas}, P.~A., {Drinkwater}, M.~J., \& {Evstigneeva}, E. 2008, \mnras, 389,
  102

\bibitem[{{van Dokkum} \& {Conroy}(2010)}]{vanDokkum2010}
{van Dokkum}, P.~G. \& {Conroy}, C. 2010, \nat, 468, 940

\bibitem[{Vanzella {et~al.}(2017)Vanzella, Calura, Meneghetti, Mercurio,
  Castellano, Caminha, Balestra, Rosati, Tozzi, De~Barros, Grazian, D'Ercole,
  Ciotti, Caputi, Grillo, Merlin, Pentericci, Fontana, Cristiani, \&
  Coe}]{Vanzella2017}
Vanzella, E., Calura, F., Meneghetti, M., {et~al.} 2017, Monthly Notices of the
  Royal Astronomical Society, 467, 4304

\bibitem[{{Weidner} {et~al.}(2004){Weidner}, {Kroupa}, \&
  {Larsen}}]{Weidner2004}
{Weidner}, C., {Kroupa}, P., \& {Larsen}, S.~S. 2004, \mnras, 350, 1503

\bibitem[{{Weidner} {et~al.}(2013){Weidner}, {Kroupa}, {Pflamm-Altenburg}, \&
  {Vazdekis}}]{Weidner2013}
{Weidner}, C., {Kroupa}, P., {Pflamm-Altenburg}, J., \& {Vazdekis}, A. 2013,
  \mnras, 436, 3309

\bibitem[{{Wright} {et~al.}(1988){Wright}, {Joseph}, {Robertson}, \&
  {Meikle}}]{Wright88}
{Wright}, G.~S., {Joseph}, R.~D., {Robertson}, N.~A. and.~{James}, P.~A., \&
  {Meikle}, W.~P.~S. 1988, \mnras, 233, 1

\bibitem[{Yan {et~al.}(2017)Yan, Jerabkova, \& Kroupa}]{Yan2017}
Yan, Z., Jerabkova, T., \& Kroupa, P. 2017, arXiv.org, arXiv:1707.04260

\bibitem[{{Zhang} \& {Bell}(2017)}]{Zhang2017}
{Zhang}, Y. \& {Bell}, E.~F. 2017, \apjl, 835, L2

\end{thebibliography}

\begin{appendix}

        \section{Additional figures and procedures}\label{app:extSED} 
        To make the text of the main paper more continuous, we present 
        a part of the figures in this Appendix.
At first we would like to present here redshift as a function of time
        from the Big-Bang. Even though this plot is elementary, it is not
        straightforward to find it in other literature in this format. We
        marked the redshift-time points that are relevant for this study.
        \begin{figure}[ht!] \begin{center}
                        \scalebox{1.0}{\includegraphics{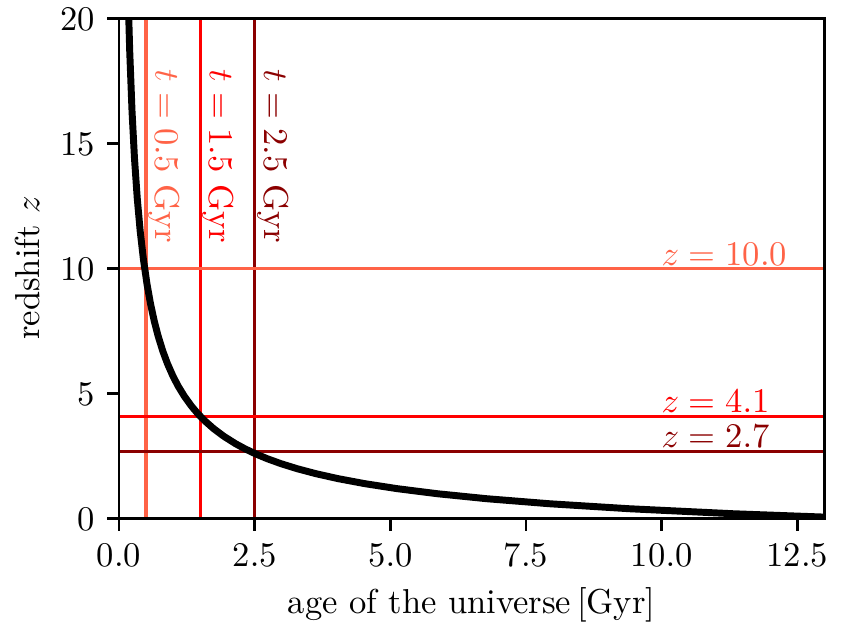}} \end{center}
                \caption{Redshift $z$ as a function of time from the Big-Bang (age of the
                        Universe) adopting for this study the standard $\Lambda$CDM
                        cosmology with Planck parameters (see Sec.~\ref{Sec:red}).} 
                \label{fig:tz}
        \end{figure}
        
        In addition, since the PEGASE SED's minimal wavelength is
        $\approx 1000\,\AA$, and because at higher redshifts ($z\gtrapprox 6$)
        this region contributes to the standard optical filters, we
        extrapolate the computed SEDs at smaller wavelengths using a
        black-body approximation. To avoid adding an additional uncertainty to our
        results, we do not use the extrapolations to make any predictions
        or conclusions, but strictly as a demonstration of an approximate
        trend.  
        
\begin{figure*}[ht!]
                \begin{center}
                        \scalebox{1.0}{\includegraphics{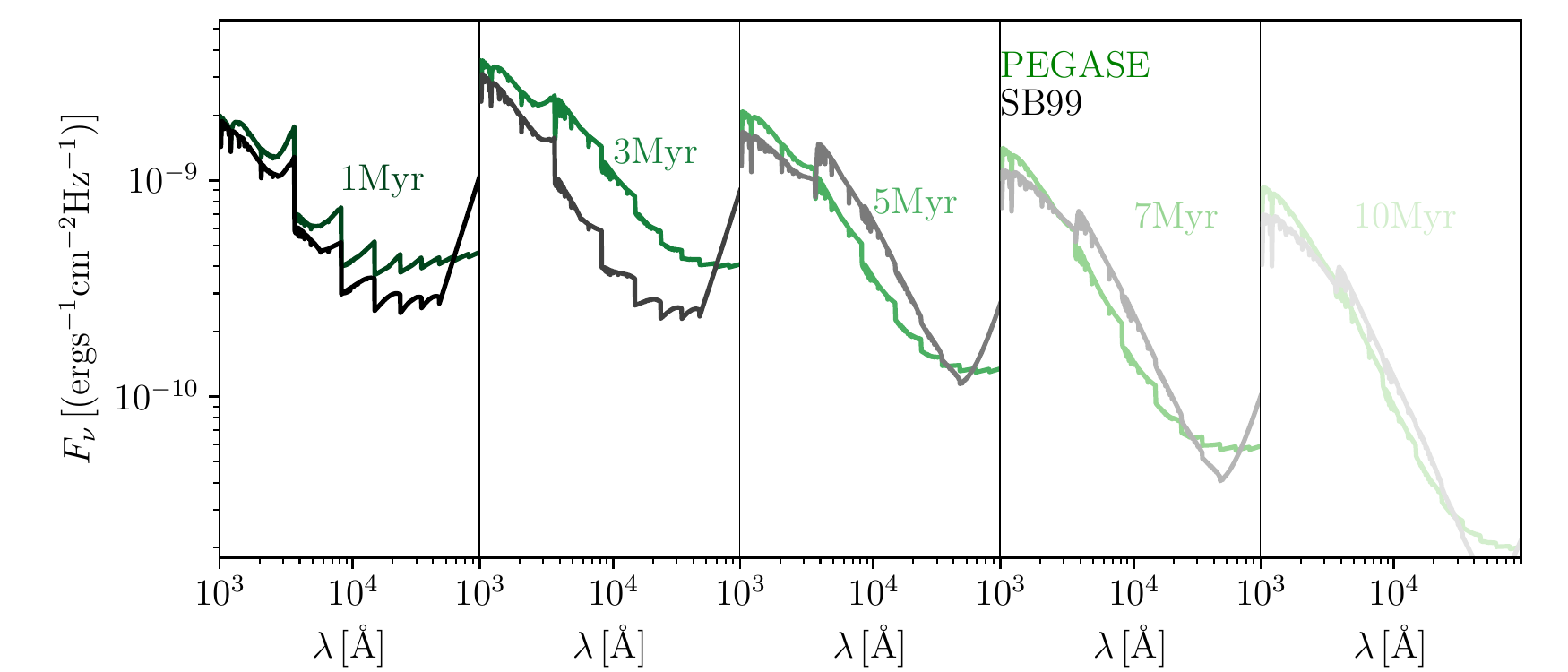}} \end{center}
                \caption{Time evolution of SEDs for a $10^9\,\mathrm{M_{\odot}}$
                        system normalized to the distance of $10\,\mathrm{pc}$ for the
                        Salpeter IMF slope $\alpha=2.35$ within the mass range (1,100)$M_{\odot}$ and
                        metallicity $Z=0.001$. We compare computed PEGASE models and
                        downloaded SB99 models with the same parameters. The SEDs are
                        normalized to the distance of 10 pc. The stellar mass
                                range used here has been adopted only for this comparison
                                for computational ease. } \label{fig:SED_SB99} \end{figure*}

        As a consistency check and also as a basic estimation of the
        difference between different stellar population codes, we computed the
        same set of SEDs with the StarBurst99--code \citep{Leitherer1999}
        online library. The results are plotted in Fig.~\ref{fig:SED_SB99}
        showing agreement in the general characteristic of the SED and in the
        time evolution. The maximum differences in the SEDs, estimated only in the region
        which is computed by PEGASE without extrapolating, reach a factor of a
        few and are expected since different stellar evolutionary tracks are
        used in both codes. These differences represent the minimal
        uncertainty that needs to be considered if our results are compared
        with observations.
        
\begin{figure*}[ht!] \begin{center}
                        \scalebox{1.0}{\includegraphics{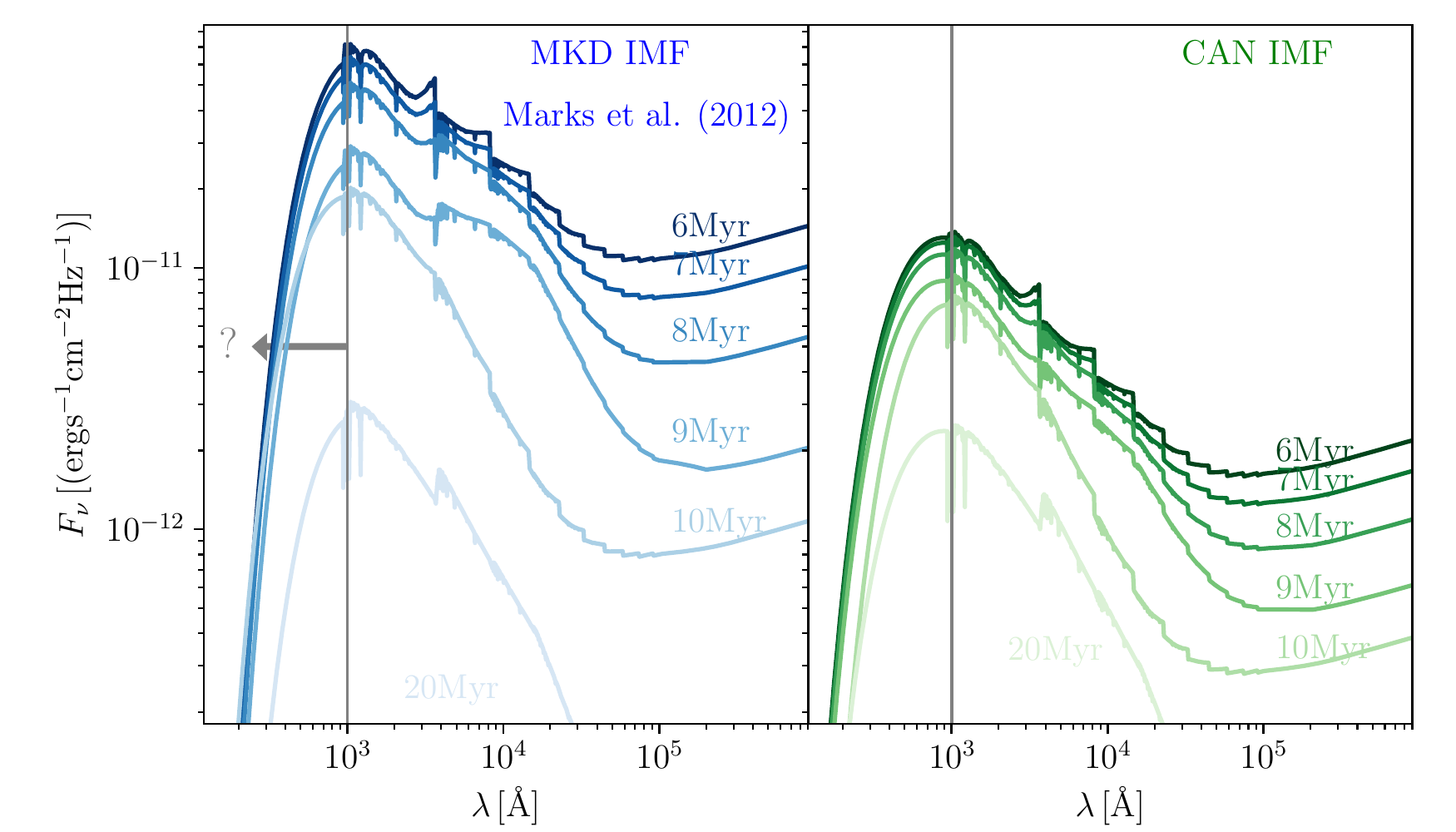}} \end{center}
                \caption{PEGASE time evolution of SEDs for [Fe/H]=-2 and a
                        representative initial stellar mass of $10^7\,\mathrm{M_{\odot}}$ as
                        if it is at a distance of~10 pc.  Since we assume that the star
                        formation lasts for 5 Myr, we start with the SED at 6 Myr to avoid
                        overlapping of lines.  {\bf Left panel:} The evolution for the MKDP
                        IMF. {\bf Right panel:} The evolution for the CAN\@ IMF.  The region
                        below a wavelength of $\approx 1000\,\AA$ (marked by arrow and
                        question mark) is the black-body approximation. The PEGASE code does
                        not predict values in this region. More details are given in the text.
                } \label{fig:MKDP_SED_T} \end{figure*}

\begin{figure*}[ht!] \begin{center}
                        \scalebox{1.0}{\includegraphics{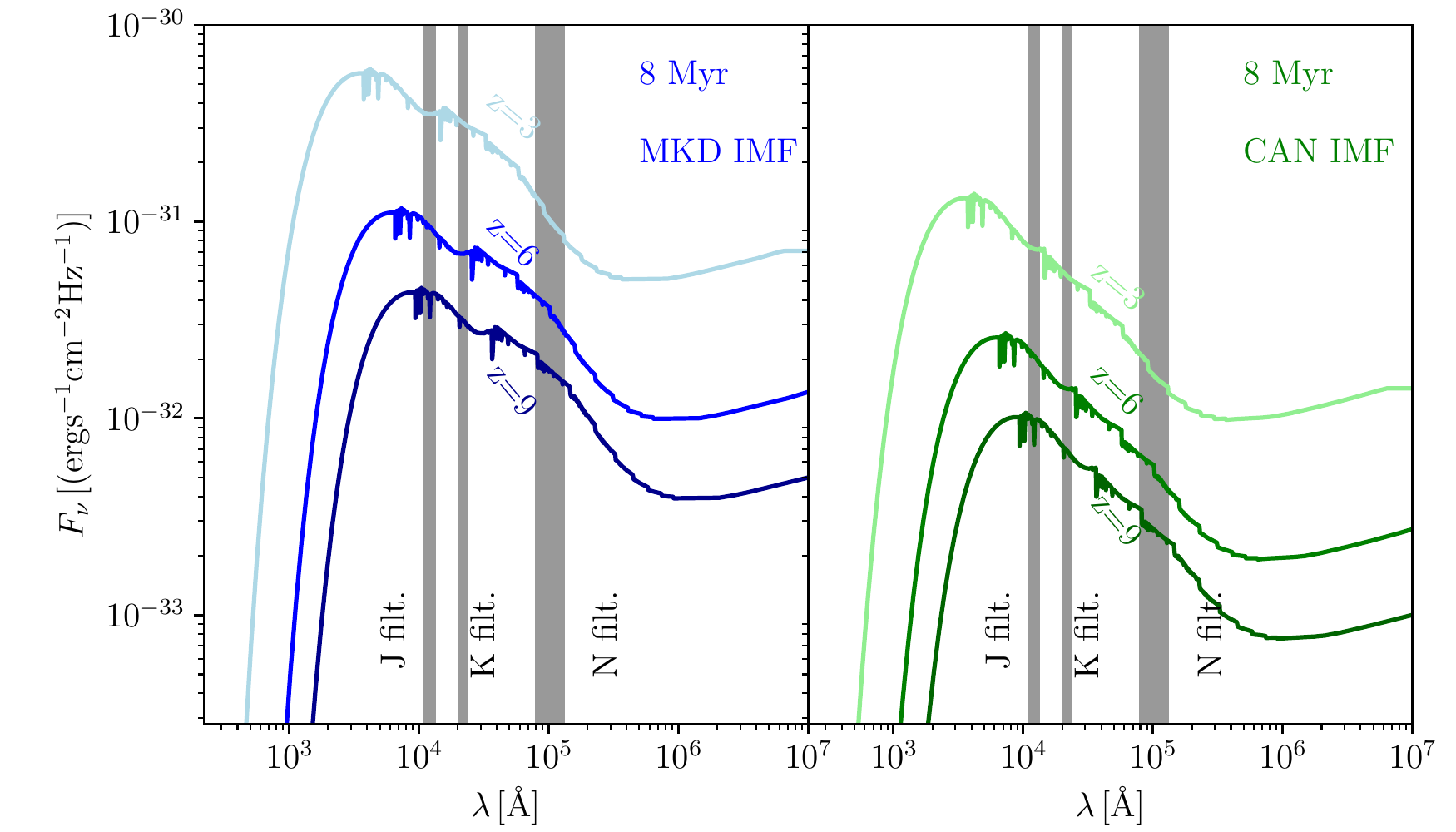}} \end{center}
                \caption{SEDs of 8-Myr-old UCDs at different
                        redshifts. For this purpose, we arbitrarily chose the SED of an 8-Myr
-old stellar population with metallicity [Fe/H]=-2 and corrected the
                        spectrum for the wavelength shift and luminosity fainting with
                        luminosity distance for redshifts 3, 6, and 9.  {\bf Left panel:}
                         SEDs of the MKDP IMF. {\bf Right panel:} SEDs of the CAN IMF.  The photometric filters, here approximated
                        by rectangular profiles as shaded vertical regions, are shown. The
                        J, K, and N filters are used in the colour analysis of the data.
                } \label{fig:SED_RED} \end{figure*}

\begin{figure*}[ht!] \begin{center}
                        \scalebox{1.0}{\includegraphics{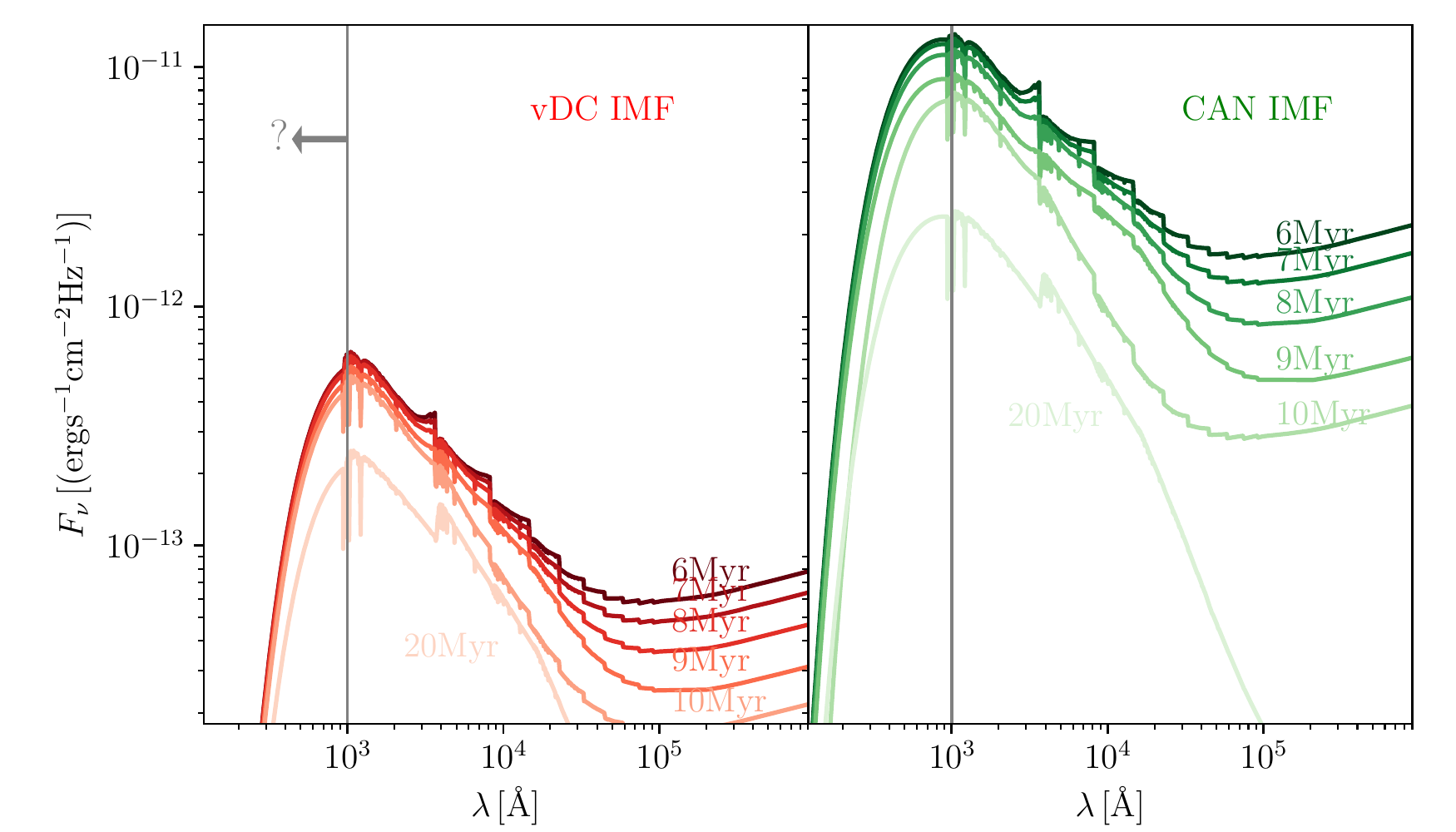}} \end{center}
                \caption{As Fig.~\ref{fig:MKDP_SED_T} but for the vDC IMF ({\bf left
                                panel}), with the CAN IMF ({\bf right panel}, identical to
                        Fig.~\ref{fig:MKDP_SED_T}) shown here as a benchmark.}
                \label{fig:VD_SED_T} \end{figure*}

\begin{figure*}[ht!] \begin{center}
                        \scalebox{1.0}{\includegraphics{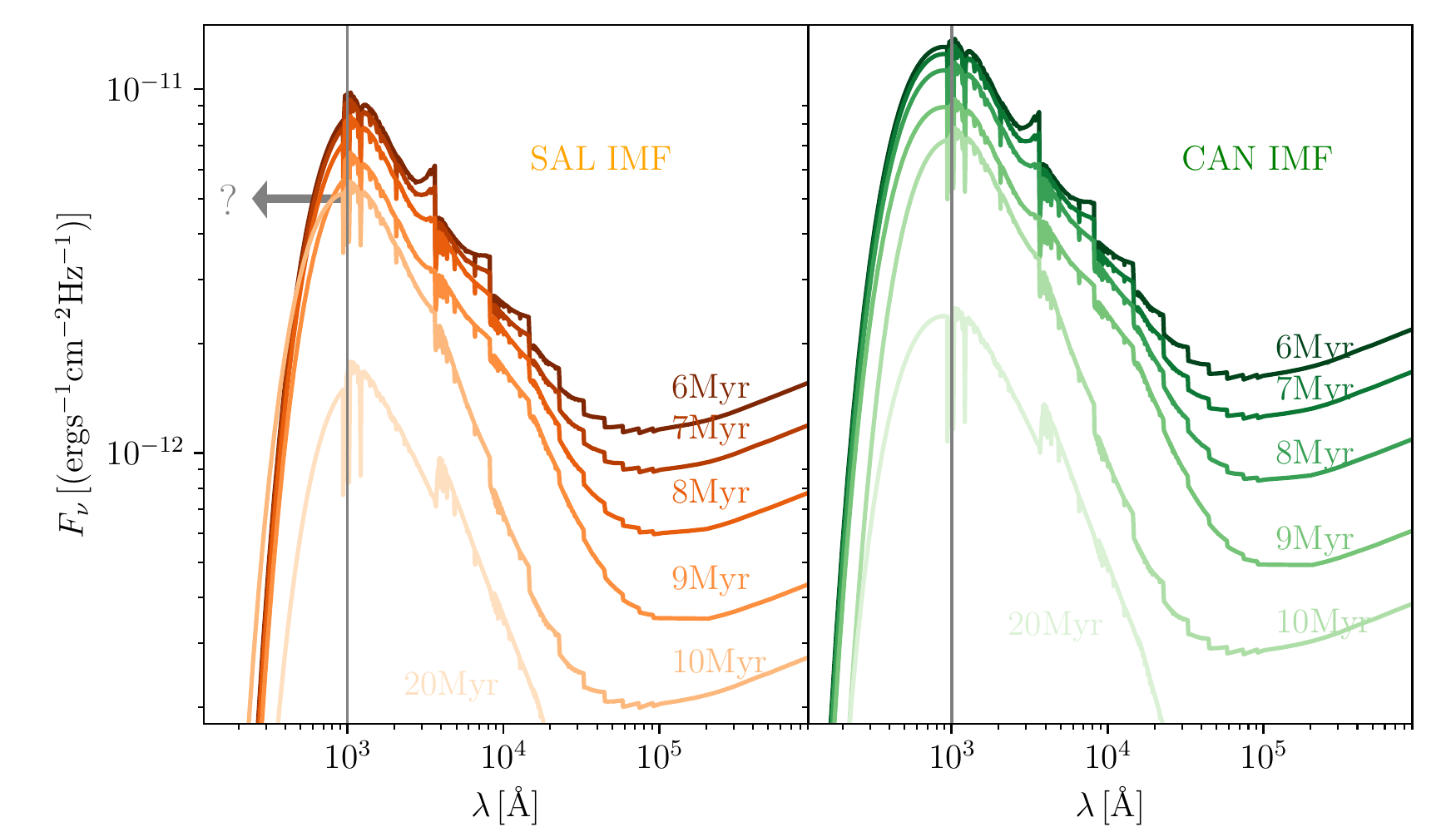}} \end{center}
                \caption{
                        As Fig.~\ref{fig:MKDP_SED_T} but for the SAL IMF ({\bf left
                                panel}), with the CAN IMF ({\bf right panel}, identical to
                        Fig.~\ref{fig:MKDP_SED_T}) shown here as a benchmark.}
                \label{fig:BH_SED_T} \end{figure*}

        The time evolution of SEDs for the \textcolor{blue}{MKDP} IMF
        is plotted in Fig.~\ref{fig:MKDP_SED_T} in comparison to the
        \textcolor{green}{CAN} IMF.  The same time evolution for the
        \textcolor{red}{vDC} and \textcolor{orange}{SAL} IMFs is plotted in
        Figs.~\ref{fig:VD_SED_T} and  \ref{fig:BH_SED_T}.  The wavelength
        shift and fainting proportional to the inverse square of luminosity
        distance with redshift is shown in Fig.~\ref{fig:SED_RED}.

        \begin{figure}[ht!] \begin{center}
                        \scalebox{1.0}{\includegraphics{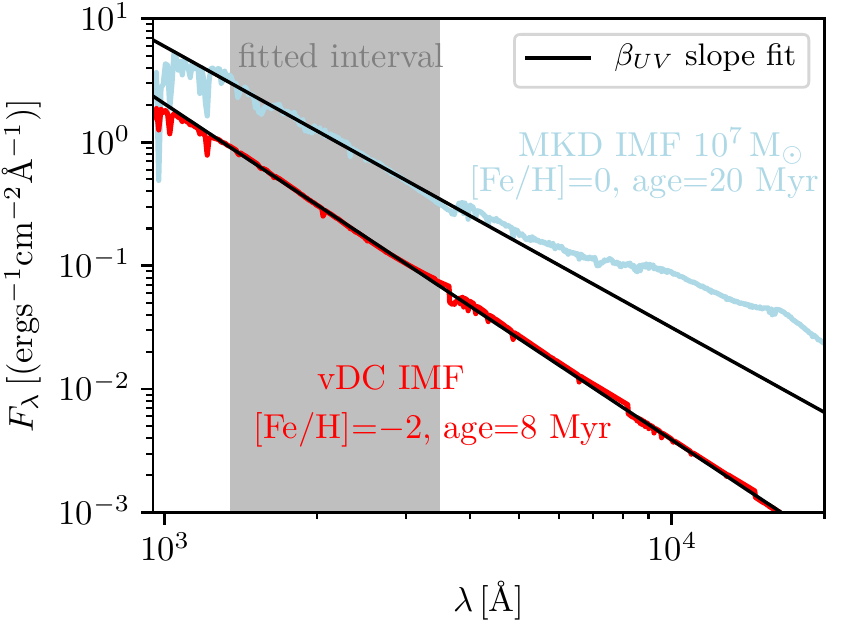}} \end{center}
                \caption{Fit to a SED in the grey shaded region shows the slope
                        $\beta_{UV}$ for two different UCDs with different age, the same
                        mass $10^7\,M_{\odot}$ , and different metallicity.  We can see that
                        in the grey region the spectra have a smooth shape and therefore it
                        is possible to fit this part by a linear function to obtain a good
                        estimate of $\beta_{UV}$. Section~\ref{sec:betaUV} gives more
                        details.} \label{fig:beta_fit} \end{figure}

        \begin{figure*}[ht!] \begin{center}
                        \scalebox{1.0}{\includegraphics{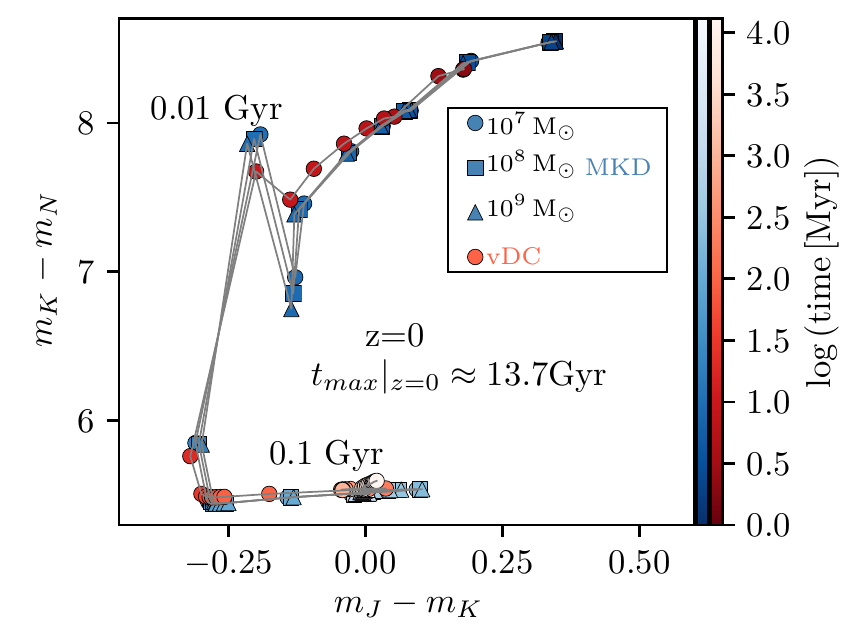}}
                        \scalebox{1.0}{\includegraphics{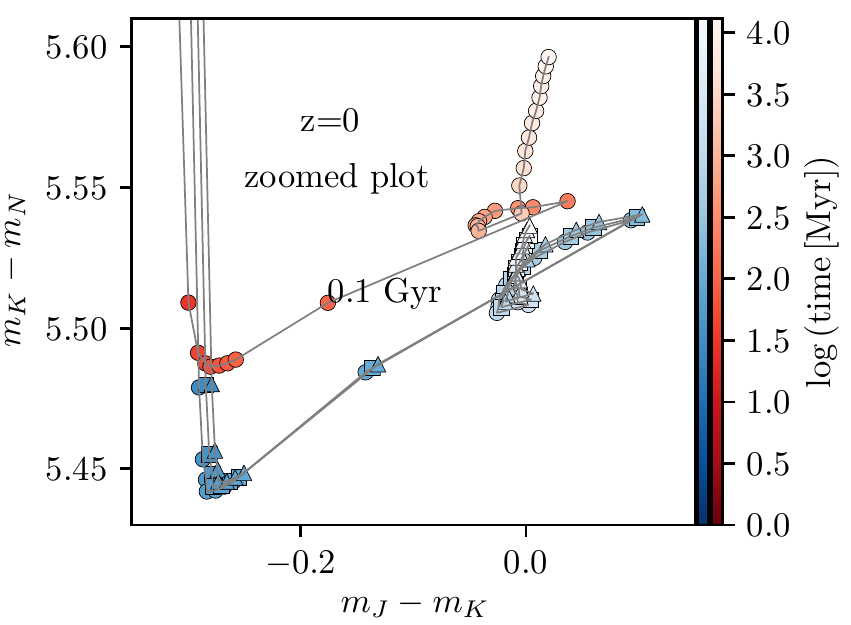}} \end{center}
                \begin{center} \scalebox{1.0}{\includegraphics{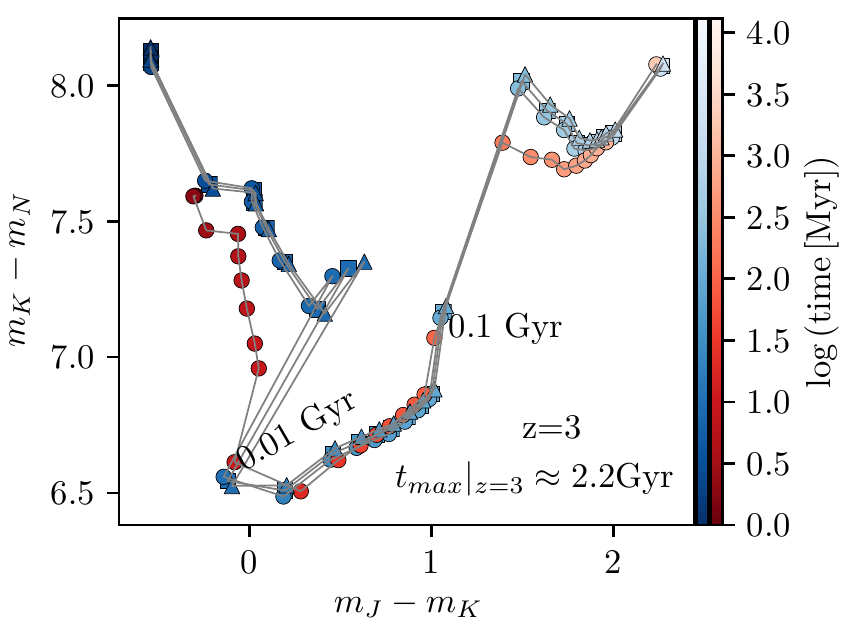}}
                        \scalebox{1.0}{\includegraphics{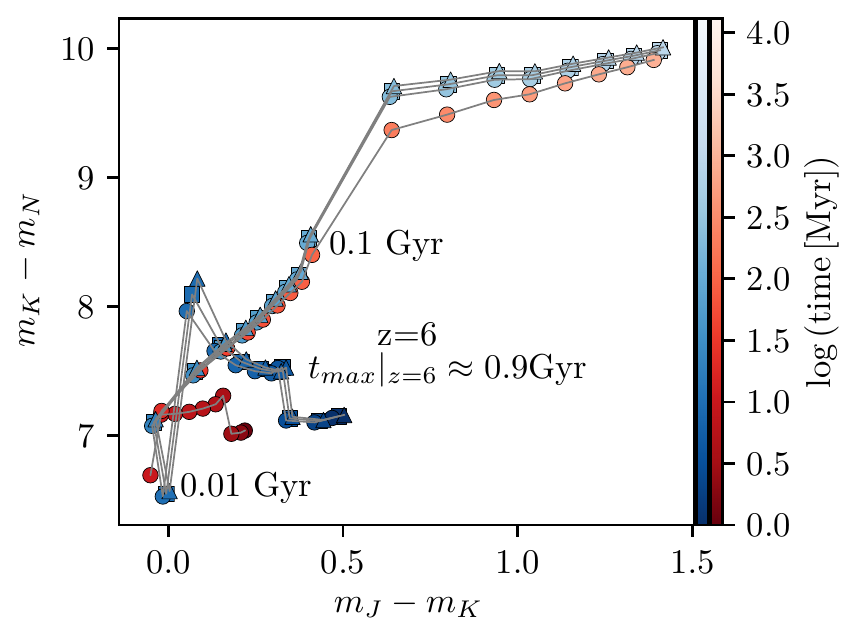}} \end{center}
                \begin{center} \scalebox{1.0}{\includegraphics{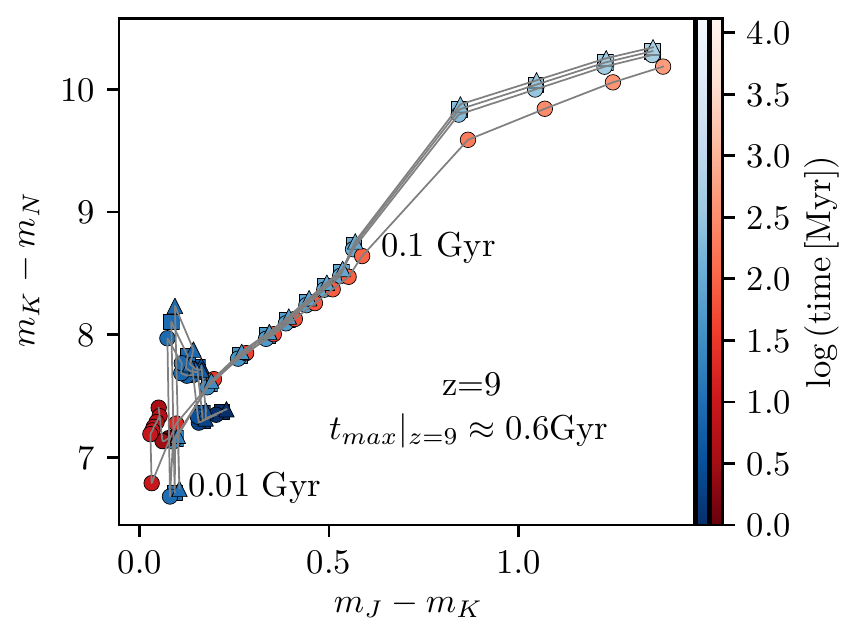}}
                        \scalebox{1.0}{\includegraphics{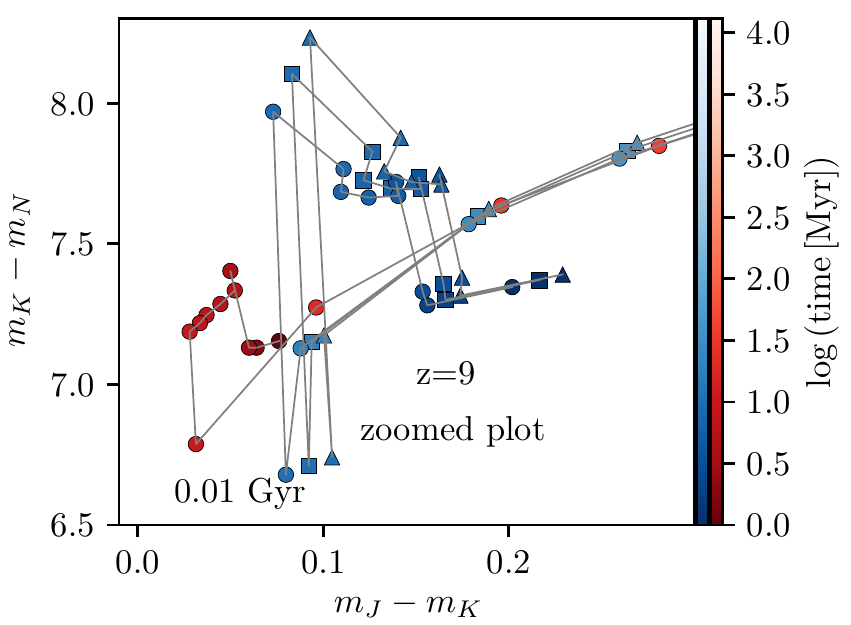}} \end{center}
                \caption{Colour-colour diagram made for standard filters J, K, and N
                        approximated here by rectangular boxes in Fig.~\ref{fig:SED_RED}
                        showing the comparison of the vDC IMF and the MKDP IMF for different
                        initial UCD masses, $M_{UCD}$.  Since according to the $\Lambda$CDM
                        cosmological model the upper limit to the age of the universe is
                        $t_{\mathrm{max}}\approx 13.7\,\mathrm{Gyr}$, we plot only ages
                        consistent with this constraint on a corresponding redshift (here
                        values are plotted at a redshift of 0, 3, 6, and 9). The age
                        evolution is shown by a colour scale that is the same for all plots.}
                \label{fig:col_col} \end{figure*}

\end{appendix}

\end{document}